%% file: main.tex
\documentclass[conference]{IEEEtran}
\IEEEoverridecommandlockouts

\usepackage{cite}
\usepackage{amsmath,amssymb,amsfonts}
\usepackage{algorithmic}
\usepackage{graphicx}
\usepackage{textcomp}
\usepackage{xcolor}
\usepackage[hyphens]{url}
\usepackage{tcolorbox}
\input{format/defs}

\def\BibTeX{{\rm B\kern-.05em{\sc i\kern-.025em b}\kern-.08em
    T\kern-.1667em\lower.7ex\hbox{E}\kern-.125emX}}
\begin{document}

\pdfpagewidth=8.5in
\pdfpageheight=11in

\newcommand{\iscasubmissionnumber}{1267}

\pagenumbering{arabic}


\title{Efficient MoE Serving in the Memory-Bound Regime: Balance Activated Experts, Not Tokens}

\author{%
  Yanpeng Yu$^{1,*}$\thanks{$^*$Equal contribution; listed in random order.}%
  \and Haiyue Ma$^{2,*}$%
  \and Krish Agarwal$^{3}$%
  \and Nicolai Oswald$^{4}$%
  \and Qijing Huang$^{4}$%
  \and Hugo Linsenmaier$^{4}$%
  \and Chunhui Mei$^{4}$%
  \and Ritchie Zhao$^{4}$%
  \and Ritika Borkar$^{4}$%
  \and Bita Darvish Rouhani$^{4}$%
  \and David Nellans$^{4}$%
  \and Ronny Krashinsky$^{4}$%
  \and Anurag Khandelwal$^{1}$ \\ \and
  \quad\quad\quad\quad\quad\space\space$^1$Yale University \quad
  $^2$Princeton University \quad
  $^3$Carnegie Mellon University \quad
  $^4$NVIDIA
}


\thispagestyle{plain}
\pagestyle{plain}

\maketitle

\input{tex/abstract}

\input{tex/introduction}
\input{tex/background}
\input{tex/motivation}
\input{tex/design}
\input{tex/implementation}
\input{tex/evaluation}
\input{tex/discussion}
\input{tex/conclusion}


\newpage
{
\bibliographystyle{IEEEtranS}
\bibliography{bib/abr-short,bib/paper}
}

\end{document}

%% file: format/defs.tex
\usepackage{xspace}
\usepackage{algorithm, algorithmic}
\usepackage{makecell}
\usepackage{tabularx}
\usepackage{tikz}
\usepackage{enumitem}
\usepackage{amsmath,amsfonts}
\usepackage{xcolor}
\usepackage{ulem}
\usepackage{amsthm, amsmath}

\def\ie{{i.e.}}

\def\namex{\textsc{metro}}
\def\name{\namex\xspace}

\usepackage{graphicx}
\usepackage[caption=false,font=footnotesize]{subfig} 

\usepackage{soul}
\usepackage{url}
\colorlet{soulgreen}{green!30}
\colorlet{soulblue}{blue!20}
\colorlet{soulred}{red!20}
\colorlet{soulyellow}{yellow!20}

\newtheorem{lemma}{Lemma}

\newcommand{\paragraphb}[1]{\vspace{0.075in}\noindent{\bf #1.}}

\usepackage{amsthm}

\theoremstyle{definition}

\def\alltoallx{{all-to-all}}
\def\alltoall{\alltoallx\xspace}

\def\allgatherx{{all-gather}}
\def\allgather{\allgatherx\xspace}

\def\rproblemx{\textsc{min-exp-routing}}
\def\rproblem{\rproblemx\xspace}

%% file: tex/abstract.tex
\begin{abstract}
Expert Parallelism (EP) permits Mixture of Experts (MoE) models to scale beyond a single GPU. To address load imbalance across GPUs in EP, existing approaches aim to balance the number of tokens each GPU processes. Surprisingly, we find that this objective \textit{degrades} performance rather than improving it when processing is memory-bound --- a common occurrence in MoE serving, especially in the decode phase. Our analysis reveals that balancing the number of tokens processed per GPU increases the number of activated experts, exacerbating memory pressure in the memory-bound regime. 

We propose \name\footnote{\textbf{M}inimum \textbf{E}xpert \textbf{T}oken \textbf{RO}uting}, a novel token-routing algorithm for high-performance expert-parallel MoE serving in the memory-bound regime that balances the number of activated experts per GPU rather than token counts. \name achieves near-optimal routing quality with minimal computational overhead by jointly optimizing algorithmic efficiency and leveraging the GPU's parallel processing power. To guarantee routing quality, \name also employs a novel allGather scheme to gather global top-$k$ knowledge, which has minimal overhead compared to conventional allToAll. Our evaluation of \name against EPLB on both real systems (vLLM over 8 A100 GPUs) and a proprietary simulator (8-16 B200 GPUs) shows that \name reduces decode latency by $11$ - $22\%$, and total token throughput by $3$ - $21\%$ for Qwen3 and DeepSeek-V3 serving, where prefill and decode phases are co-deployed. In addition, by trading latency headroom for throughput, \name improves decode throughput by up to $4.11\times$ over EPLB at a fixed decode SLO.

\end{abstract}


%% file: tex/introduction.tex
\section{Introduction}
\label{sec:intro}

Mixture-of-Expert (MoE) models~\cite{qwen3technicalreport, deepseekai2025deepseekv3technicalreport} are growing increasingly popular due to their lower computational cost per token generation compared to denser models. When MoE models' footprint scales beyond a single GPU's memory capacity, Expert Parallelism (EP)~\cite{fedus2022switchtransformersscalingtrillion, lepikhin2020gshardscalinggiantmodels, shazeer2017outrageouslylargeneuralnetworks} is emerging as the de facto choice for scaling MoE models to multiple GPUs. In it, experts are placed on different GPUs and process tokens in parallel. EP is often augmented with load-balancing mechanisms~\cite{eplb-repo, moesys, flexmoe, fastermoe, smartmoe, cheng2025barbariansgateaiupending, wei2023prophet, rajbhandari2022deepspeedmoe, hwang2023tutel, li2023lina, yao2024interlayeraffinity} to address load imbalance across GPUs. This includes creating replicas of popular experts (expert replication), strategically placing these replicas across GPUs (expert placement), and routing tokens across the replicas of an expert (token routing\footnote{\label{fn:token_routing}We refer to token routing as the process of selecting specific expert replica(s) to `activate' to serve a particular token, \textit{not} the process of choosing the top-$k$ experts.}).

The underlying principle of the load-balancing mechanisms outlined above is minimizing the runtime of the slowest GPU in the pool, since it determines the end-to-end runtime. Existing EP expert placement/replication and token-routing algorithms~\cite{eplb-repo, moesys, flexmoe, fastermoe, smartmoe, cheng2025barbariansgateaiupending, wei2023prophet, rajbhandari2022deepspeedmoe, hwang2023tutel, li2023lina, yao2024interlayeraffinity} attempt to achieve this by balancing the number of tokens processed by each GPU (\ie, EP rank). Such ``token-balancing'' implicitly assumes that the GPU runtime for inference requests scales linearly with the number of tokens processed. However, this is only true when the inference workload is \textit{compute-bound}, \ie, the GPU runtime is bottlenecked by tensor core throughput rather than memory bandwidth~(\S\ref{ssec:mem_bound}).


The reality, however, is more nuanced --- real-world serving workloads comprise a mix of \textit{both} compute-bound (\ie, prefill) and memory-bound phases (\ie, decode). Indeed, we find that in the memory-bound regime, a GPU's MoE layer runtime is not governed by the number of processed tokens, but instead determined by the number of `activated' expert replicas --- the replicas that actually process tokens in a given batch. This is because in the memory-bound regime, runtime increases with the amount of memory traffic, which is dominated by loading expert weights from GPU memory to tensor cores.
Unfortunately, existing EP load-balancing schemes that balance tokens across GPUs inadvertently increase the number of activated expert replicas, \textit{degrading} performance in the memory-bound regime~(\S\ref{ssec:balance_experts}).

\begin{figure}[t!]
  \centering
 {\includegraphics[width=0.5\textwidth]{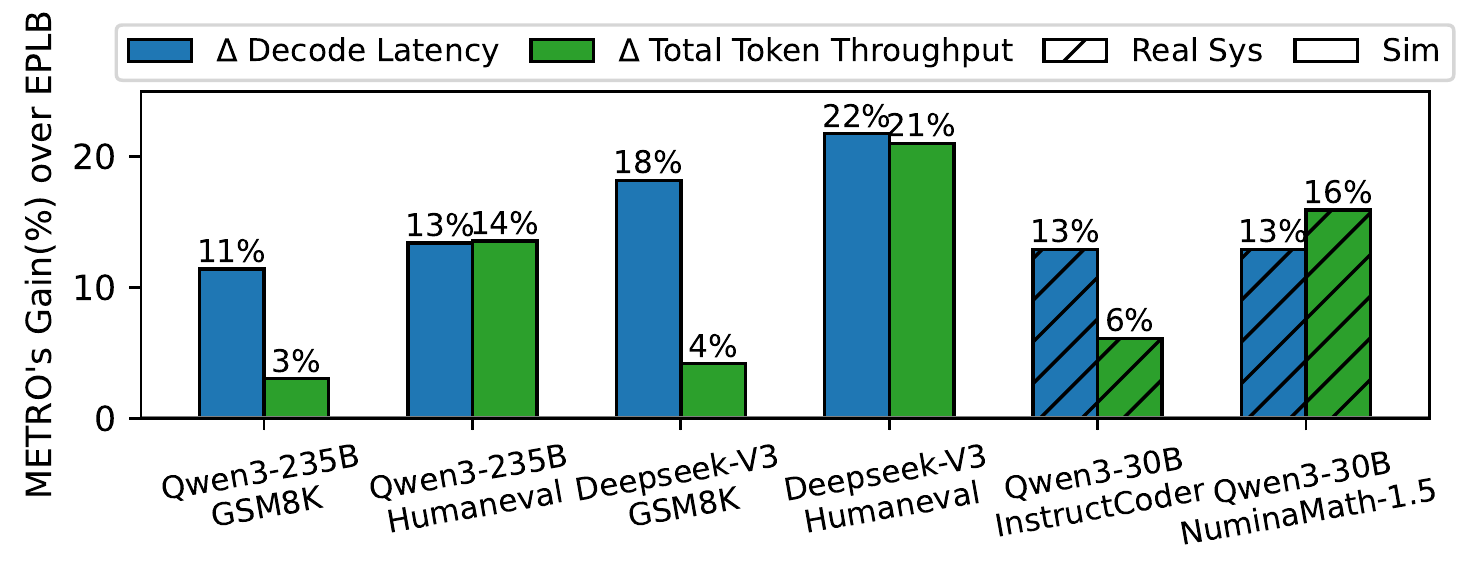}}
  \hfill
  \caption{\textbf{\name achieves universal (up to $22\%$) performance improvement on both decode latency and total token throughput (prefill-decode co-deployed) over EPLB's token routing across models, datasets, and hardware setups. Results are from both real system evaluation and simulation (\S\ref{ssec:endtoend_perf}). Replication ratio: $50\%$. Placement algorithm: EPLB.}}\label{fig:figure_1}
\end{figure}

These observations motivate our central research question: in a MoE serving system that comprises both compute-bound prefill and memory-bound decode phases, \textit{can we balance EP load without hurting --- and ideally improving --- the performance of the memory-bound phase (\ie, decode) performance, so that both phases can benefit from replication strategies?}
Since tailoring expert replication and placement to the decode phase can interfere with the prefill phase's performance, we restrict our focus to the token-routing algorithm in this work.

We present \name, a novel token-routing algorithm for EP load balancing in the memory-bound regime that minimizes GPU runtime by \textit{minimizing the number of activated expert replicas across GPUs}.

Realizing \name requires addressing several distinct challenges (\S\ref{sec:design}). First, we find that an optimal solution to the routing problem --- \ie, minimizing the number of activated expert replicas across GPUs --- incurs prohibitive computational overheads in practice. \name therefore adopts a GPU-native greedy approximate algorithm that preserves \textit{near-optimal} routing quality while limiting computational overheads. Second, unlike balancing tokens, minimizing activated experts in \name requires each GPU to collect global top-$k$ expert selections from all GPUs to make informed routing decisions. To enable this, \name replaces the conventional \alltoall EP dispatch with a novel, low-overhead \allgather token dispatching scheme that disseminates the global top-$k$ information across GPUs.


We have evaluated \name on both real systems (vLLM~\cite{vllm} over 8 NVIDIA A100 GPUs~\cite{nvidia_a100_whitepaper_2020}) and a proprietary industrial multi-GPU performance simulator (8–16 B200 GPUs~\cite{nvidia_blackwell_tech_overview_2025}). Our results show that \name consistently outperforms the baseline EPLB routing algorithm across various settings and replication ratios. Specifically, as shown in Fig.~\ref{fig:figure_1}, \name achieves up to $21.8\%$ reduction in decode latency at $1.5\times$ replication, translating into up to $21.0\%$ improvement in total token throughput by effectively minimizing the number of activated experts~(\S\ref{ssec:endtoend_perf}). Moreover, \name achieves up to $4.11\times$ higher decode throughput compared to EPLB with a fixed decode SLO, by trading its latency headroom for throughput~(\S\ref{ssec:decode_only_analysis}).
These gains are observed across a wide range of real-world MoE models (Qwen3-30B, Qwen3-235B~\cite{qwen3technicalreport}, and DeepSeek-V3~\cite{deepseekai2025deepseekv3technicalreport}) and workloads (InstructCoder~\cite{li2024instructcoderinstructiontuninglarge}, NuminaMath-1.5~\cite{numina_math_datasets}, GSM8K~\cite{cobbe2021gsm8k}, and Humaneval~\cite{chen2021evaluating}).


In summary, this paper makes the following contributions:
\begin{itemize}
    \item We identify that in the memory-bound regime, a GPU's MoE layer runtime is determined by the number of `activated' expert replicas.
    \item We show that, in the memory-bound regime, existing EP load-balancing schemes that balance tokens degrade performance by inflating the number of activated experts.
    \item We present \name, a novel token-routing algorithm for EP load balancing in the memory-bound regime, whose key idea is to minimize GPU runtime by minimizing the number of activated expert replicas.
    \item We implement and evaluate \name in both a real-world vLLM-based system and a proprietary industrial simulator. Our results show that \name reduces MoE decode latency while improving total token throughput compared to existing EP load-balancing schemes.
\end{itemize}

%% file: tex/background.tex
\section{Background}
\label{sec:background}

\subsection{Autoregressive LLM Inference}
LLM inference consists of two phases: a prefill phase followed by a decode phase. In prefill, the model processes the entire input prompt in a single forward pass to produce the first output token. Since this phase processes many tokens at once, it is typically compute-bound and achieves high GPU utilization. In the subsequent decode phase, the model generates tokens autoregressively, feeding each newly generated token back into the model until an end-of-sequence token is produced. Since the decode phase processes only one token per request, even when batching across multiple requests, it cannot fully saturate the GPU cores and is typically memory-bound.
Due to such heterogeneity, recent works have proposed disaggregating prefill and decode for large-scale deployments with multiple server nodes and up to hundreds of GPUs~\cite{zhong2024distserve, pratyush2024splitwise, nvidia2025dynamo, qin2025mooncake}. However, on smaller systems --- e.g., a single server with a few GPUs --- prefill and decode are still commonly co-deployed~\cite{agrawal2024sarathi_serve, fu2024serverlessllm, xupeng2024specinfer, gao2024cachedattention, aditya2025podattention, zhu2025nanoflow}, due to various reasons as we will discuss in \S\ref{sec:discussion}. In this paper, we first target the co-deployed setting for small multi-GPU systems and later discuss how our techniques also benefit disaggregated prefill–decode deployments in \S\ref{sec:discussion}.

\begin{figure}[t!]
  \centering
 \includegraphics[width=0.475\textwidth]{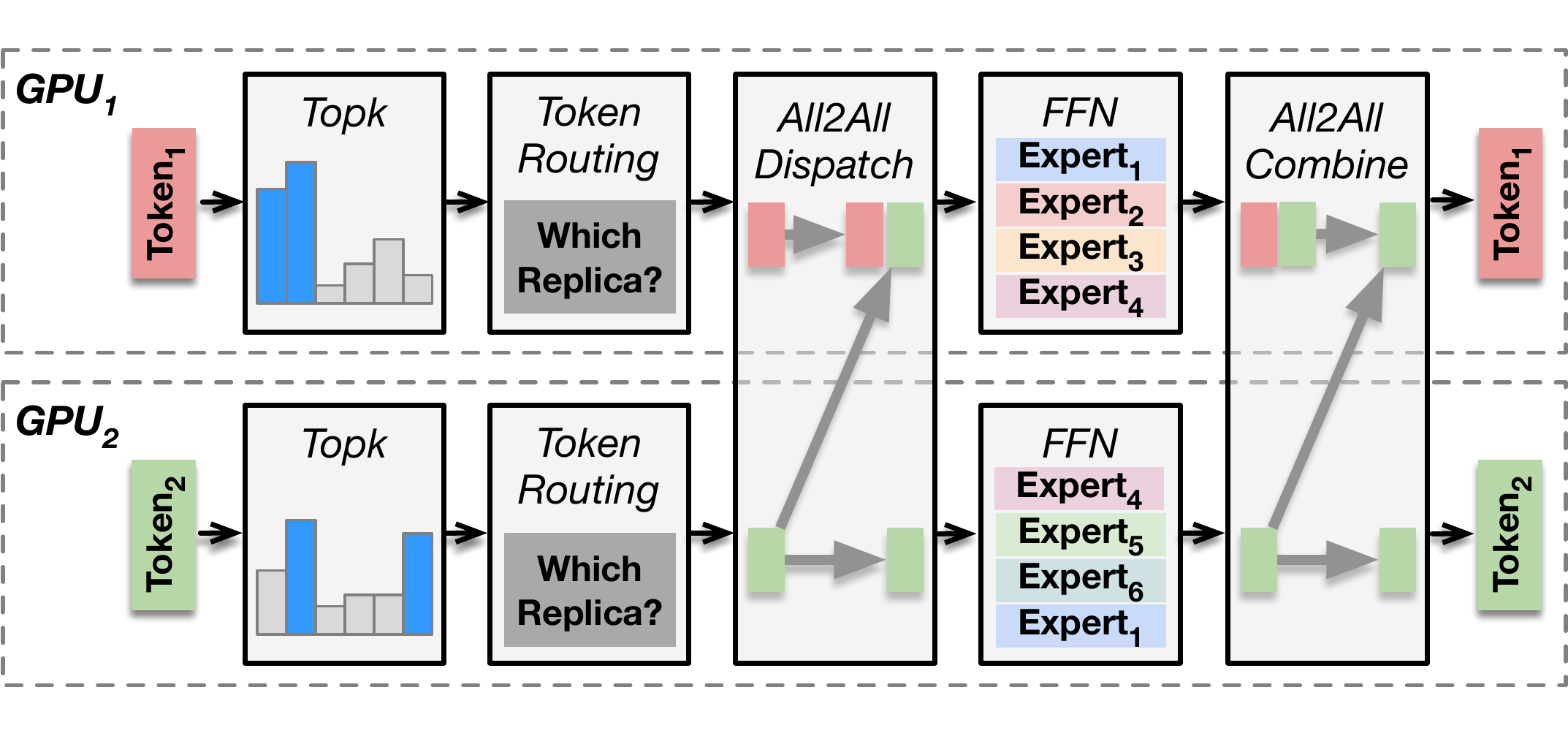}\label{fig:background_moe_ep}
  \hfill
  \caption{\textbf{Expert-parallel MoE inference workflow with expert placement and replication, as well as token routing -- the algorithm to dynamically route tokens to expert replicas.}} 
\end{figure}

\subsection{MoE Inference and Expert Parallelism}
Mixture-of-Expert (MoE) models~\cite{qwen3technicalreport, deepseekai2025deepseekv3technicalreport} are growing increasingly popular due to their lower computational cost per token generation compared to dense models. When MoE models' footprint scales beyond a single GPU's memory capacity, Expert Parallelism (EP)~\cite{fedus2022switchtransformersscalingtrillion, lepikhin2020gshardscalinggiantmodels, shazeer2017outrageouslylargeneuralnetworks} is arguably the most popular algorithm to scale MoE to multiple GPUs. In it, experts are placed on different GPUs and process tokens in parallel. EP is often augmented with systematic load-balancing mechanisms~\cite{eplb-repo, moesys, flexmoe, fastermoe, smartmoe, cheng2025barbariansgateaiupending, wei2023prophet, rajbhandari2022deepspeedmoe, hwang2023tutel, li2023lina, yao2024interlayeraffinity}. This includes creating replicas of popular experts (expert replication), strategically placing these replicas across GPUs (expert placement), and routing tokens across the replicas of an expert after top-$k$ selection (token routing).

Fig.~\ref{fig:background_moe_ep} shows the expert-parallel inference workflow for MoE's expert layer, augmented with the mechanisms described above. Each GPU first computes a top-$k$ value for each of its input tokens, which determines the set of experts to activate for that token. Then, token routing determines which physical expert replica each token should be routed to. Next, an \alltoall communication is used to dispatch tokens to the GPUs that host the identified target expert replica. Then, tokens' output embeddings are computed using the corresponding expert replicas. Finally, each token's output embedding is sent back to its original GPU using another \alltoall communication, and combined there to produce the final output.

\subsection{Existing replication and routing algorithms balance tokens}


To our knowledge, all existing EP expert placement/replication and token-routing algorithms aim to balance the number of tokens processed across GPUs (\ie, EP ranks). For example, DeepSeek's Expert Parallelism Load Balancer (EPLB)~\cite{eplb-repo}, the state-of-the-art expert placement and replication algorithm, balances tokens in two steps. First, it creates replicas for each expert, with replica count proportional to the number of tokens processed by the expert in the last time window. Second, it places these replicas on GPUs to balance the number of tokens they are expected to process in the next time window. EPLB's replica placement algorithm (i.e., the second step) assumes that the token-routing algorithm distributes each expert's token evenly across its replicas. Most EPLB implementations (e.g., by vLLM~\cite{vllm} and SGLang~\cite{sglang}) follow such a token-routing algorithm.
While other studies~\cite{moesys, flexmoe, fastermoe, smartmoe, cheng2025barbariansgateaiupending, wei2023prophet, rajbhandari2022deepspeedmoe, hwang2023tutel, li2023lina, yao2024interlayeraffinity} have proposed various other expert placement, replication, and token-routing algorithms, they share the same optimization objective as EPLB --- balancing tokens across GPUs.


%% file: tex/motivation.tex
\section{Motivation}
\label{sec:motivation}

Contrary to conventional wisdom, we find that balancing tokens can degrade, rather than improve, MoE inference performance across a large class of real-world settings. Specifically, we first show that balancing tokens implicitly assumes the compute-bound regime, whereas MoE inference comprises of compute-bound (\ie, prefill) and memory-bound (\ie, decode) phases~(\S\ref{ssec:mem_bound}). We then show that in the memory-bound regime, balancing tokens \textit{degrades} performance rather than improving it, by increasing the number of activated experts and exacerbating the memory pressure~(\S\ref{ssec:balance_experts}).

\subsection{Balancing tokens assumes the compute-bound regime while MoE's decode phase is increasingly memory-bound}\label{ssec:mem_bound}
Balancing tokens to minimize runtime assumes that the inference workload is entirely compute-bound, i.e., the GPU runtime is bottlenecked by tensor core throughput rather than memory bandwidth. This is because only in the compute-bound regime is the number of tokens processed by a GPU proportional to its runtime, making balancing tokens equivalent to balancing (and thus minimizing) GPU runtime.

\begin{figure}[t!]
  \centering
  \includegraphics[width=0.235\textwidth]{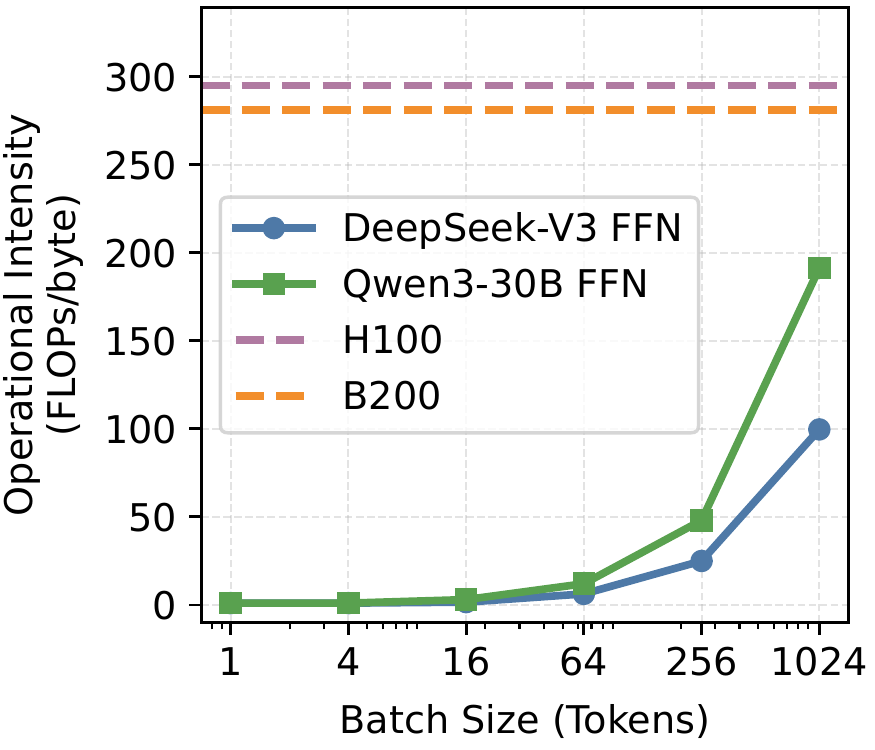}
  \hfill
  \caption{\textbf{DeepSeek-V3 and Qwen3-30B attainable operational intensities VS. FLOPs/byte ratio of H100 and B200~(\S\ref{ssec:mem_bound}).} The former is two orders of magnitude lower than the latter with batch size smaller than $64$ tokens, and $47\%$ - $3.0\times$ lower with a batch size of $1024$ tokens.}\label{fig:mtv_moe_vs_gpu_oi}
\end{figure}

While such compute-bound assumption holds in many settings --- especially those where existing token-based load balancing algorithms were originally proposed (\ie, training and prefill) --- recent studies~\cite{davies2025efficientllminferencebandwidth, yuan2024llminferenceunveiledsurvey, agrawal2024sarathi_serve} have observed that MoE inference workloads are increasingly memory-bound --- especially in the decode phase --- due to several emerging trends. These include limited inter-token weight reuse compared to dense models due to increasing MoE sparsity~\cite{moe_survey1}, the use of smaller batch sizes due to the memory capacity pressure of increasing context lengths~\cite{vllm, gemini15report, anthropic_sonnet4_1m_2025}, and the increasing FLOPs/byte ratio of recent GPU hardware~\cite{zhang2025tensorcoresbenefitmemorybound, yuan2024llminferenceunveiledsurvey}. Indeed, Fig.~\ref{fig:mtv_moe_vs_gpu_oi} shows that the attainable operational intensities (\ie, the ratio of the amount of arithmetic operations performed to the amount of data moved to/from memory~\cite{Williams2009Roofline}) during decode (following the calculation in ~\cite{davies2025efficientllminferencebandwidth}) for two state-of-the-art MoE models, DeepSeek-V3~\cite{deepseekai2025deepseekv3technicalreport} and Qwen3-30B~\cite{qwen3technicalreport}, are \textit{two orders of magnitude} lower than the FLOPs/byte ratio of recent NVIDIA datacenter GPUs (\ie, H100~\cite{nvidia_h100_whitepaper_2022} and B200~\cite{nvidia_blackwell_tech_overview_2025}) with a batch size smaller than $64$ tokens. Even batching as high as $1024$ decode tokens, the GPUs' FLOPs/byte ratio is still $47\%$ - $3.0\times$ higher than the operational intensities of the MoE models. Moreover, the models' attainable operational intensities in practice are often lower than the ideal value due to buffer capacity constraints across the memory hierarchy~\cite{huang2024mindthegap}. Therefore, it is critical to re-examine EP load balancing in the memory-bound regime, where balancing tokens does not directly translate into balancing GPU runtimes.

\subsection{Balancing tokens inflates activated experts and thus degrades performance in the memory-bound regime}\label{ssec:balance_experts}
\begin{figure}[t!]
  \centering
  \subfloat[]{\includegraphics[width=0.245\textwidth]{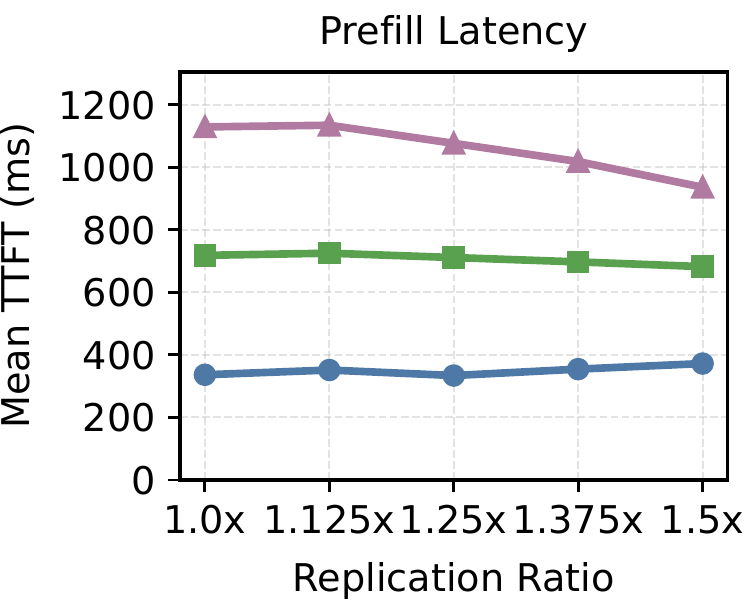}\label{fig:mtv_vllm_eplb_prefill}}
  \subfloat[]{\includegraphics[width=0.245\textwidth]{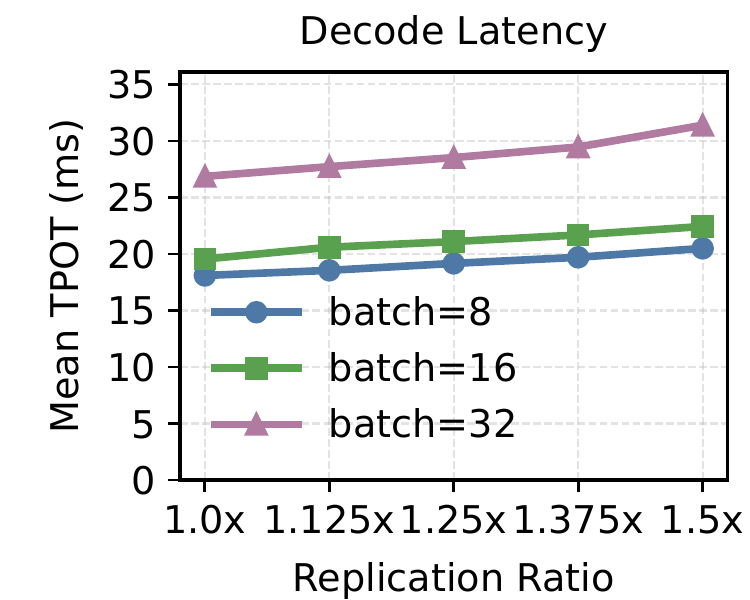}\label{fig:mtv_vllm_eplb_decode}}
  \hfill
  \subfloat[]{\includegraphics[width=0.245\textwidth]{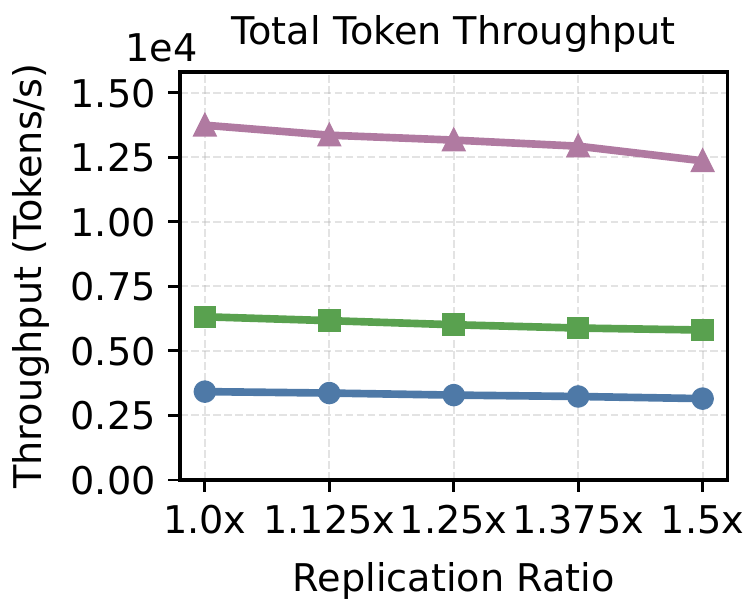}\label{fig:mtv_vllm_eplb_throughput}}
    \subfloat[]{\includegraphics[width=0.245\textwidth]{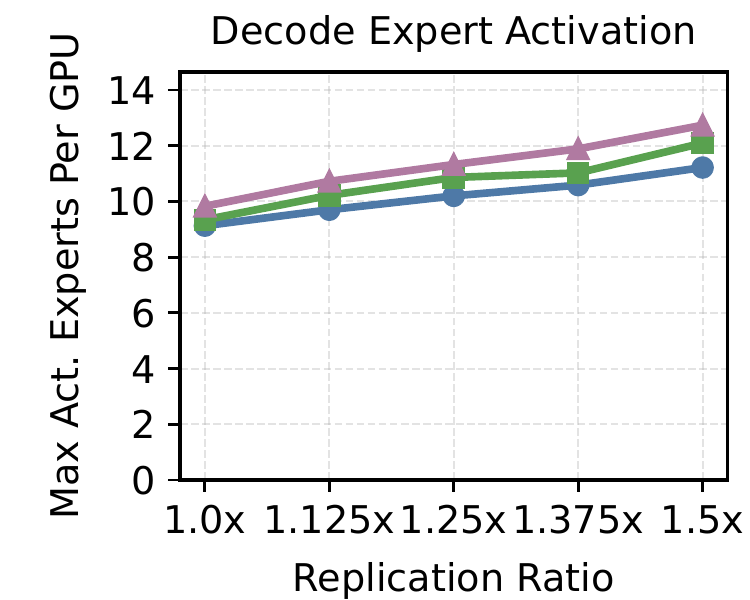}\label{fig:mtv_vllm_eplb_expert}}
  \caption{\textbf{The performance impact of EPLB on prefill latency (a), decode latency (b), overall token throughput (c), and maximum number of activated experts across GPUs per decode batch (d) for Qwen3-30B~\cite{qwen3technicalreport} on vLLM~\cite{vllm}~(\S\ref{ssec:balance_experts}).} Context length: $8K$. Dataset: InstructCoder~\cite{li2024instructcoderinstructiontuninglarge}. EPLB reduces prefill latency by $17\%$ with batch size $32$, but inflates the number of activated experts by $30\%$ with $1.5x$ replication. As a result, the decode latency increases by $14\%$ and the overall token throughput decreases by $10\%$ with $1.5x$ replication. We also find that the prefill phase can be in the memory-bound regime when the batch size is small (\ie, $8$ and $16$ requests per GPU), where EPLB cannot improve its performance.}
  \label{fig:mtv_vllm_eplb}
\end{figure}

We find that in the memory-bound regime, a GPU's MoE layer runtime is not governed by the number of processed tokens, but instead determined by the number of `activated' expert replicas --- the replicas that actually process tokens in a given batch. This is because, in the memory-bound regime, runtime scales with total memory traffic, which is dominated by loading expert weights from GPU memory to tensor cores. Indeed, the token activations' footprint is significantly smaller: the activation's memory traffic is $<0.6\%$ of the expert weights' memory traffic for a decode batch as large as $1$K tokens based on the analytical model proposed in recent work~\cite{davies2025efficientllminferencebandwidth}. We empirically confirm this observation in Fig.~\ref{fig:mtv_vllm_eplb_expert} and Fig.~\ref{fig:mtv_vllm_eplb_decode}, which shows a strong correlation between the number of activated experts and vLLM decode latency (evaluation setup detailed in \S\ref{ssec:system_setup}).

\begin{figure}[t!]
  \centering
 \includegraphics[width=0.475\textwidth]{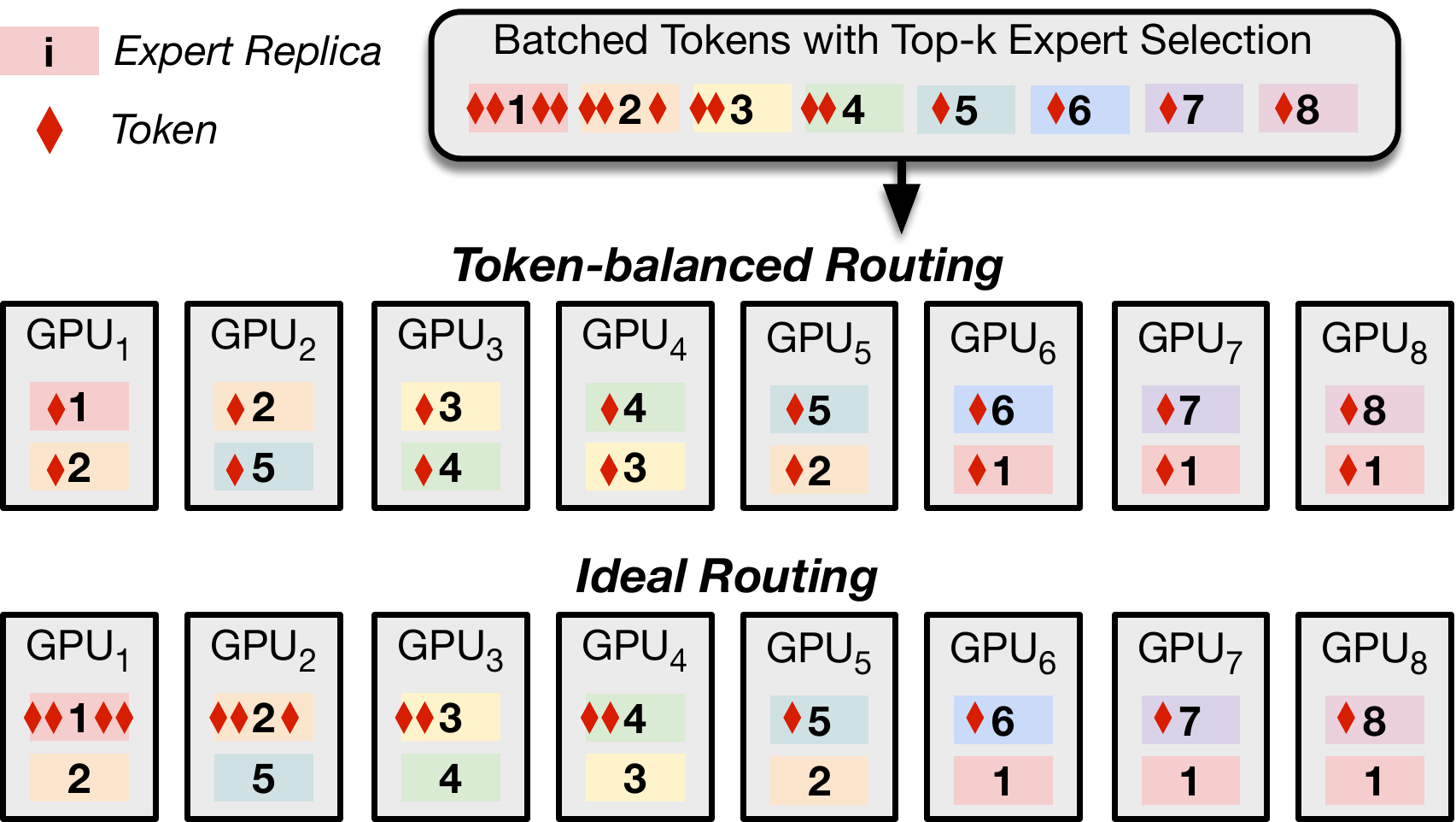}
  \hfill
  \caption{\textbf{Example: balancing tokens for token routing doubles the activated experts per GPU and the theoretical runtime in the memory-bound regime, compared to the hypothetical ideal token routing~(\S\ref{ssec:balance_experts}).}}
  \label{fig:mtv_token_balanced_vs_ideal_routing}
\end{figure}

Unfortunately, balancing tokens across GPUs increases the number of activated expert replicas, thereby degrading performance in the memory-bound regime. Fig.~\ref{fig:mtv_token_balanced_vs_ideal_routing} uses a toy example to compare two token routing schemes under the same expert replication and placement scheme for a batch of size 16. For the token-balanced routing scheme to achieve perfect token balance (\ie, two tokens per GPU), it must evenly distribute tokens across all available replicas. This causes two expert replicas to be activated on each GPU. However, as demonstrated by the hypothetical ideal routing scheme, all tokens routed to a particular expert should ideally be routed to a single replica to minimize the activated experts per GPU (\ie, one per GPU). Since runtime increases with the number of activated experts rather than the number of tokens per GPU in the memory-bound regime, balancing tokens would double the GPU runtime relative to the ideal in this example!

To understand the end-to-end performance degradation of balancing tokens in the memory-bound regime, we measure the performance impact of EPLB on prefill latency (\ie, Time-to-first-token, or TTFT), decode latency (\ie, Time-per-output-token, or TPOT), and the total token throughput for a state-of-the-art serving system, vLLM~\cite{vllm}. We use a state-of-the-art MoE model, Qwen3-30B~\cite{qwen3technicalreport} with a representative coding dataset, InstructCoder~\cite{li2024instructcoderinstructiontuninglarge}. We observe similar trends for other datasets (with results deferred to \S\ref{ssec:system_setup}). Fig.~\ref{fig:mtv_vllm_eplb} shows that while EPLB can reduce prefill latency (Fig.~\ref{fig:mtv_vllm_eplb_prefill}) by $17\%$ with up to $1.5\times$ replication when the batch size is large enough for prefill to enter the compute-bound regime (\ie, $32$ requests per GPU), it increases the number of activated experts per decode batch (Fig.~\ref{fig:mtv_vllm_eplb_expert}) as the replication factor increases --- nearly $30\%$ more activated experts per decode batch with $1.5\times$ replication. Consequently, not only does the decode latency increase with more replication (\ie, by $14\%$ with $1.5\times$ replication, Fig.~\ref{fig:mtv_vllm_eplb_decode}), but the overall token throughput also decreases (\ie, by $10\%$ with $1.5\times$ replication, Fig.~\ref{fig:mtv_vllm_eplb_throughput}).

These observations motivate our central research question: In a MoE serving system where the compute-bound prefill and memory-bound decode phases are co-deployed, \textit{can we balance EP load without hurting --- and ideally improving --- the memory-bound decode phase performance, allowing both phases to benefit from additional memory capacity?}
Since tailoring expert replication and placement to the decode phase may interfere with prefill phase performance, we restrict our focus to the token-routing algorithm.
To this end, we present \name, a novel token-routing algorithm for EP load balancing in the memory-bound regime. \name minimizes GPU runtime by \textit{minimizing the number of activated experts across GPUs}, as we detail in the next section.

%% file: tex/design.tex
\section{\name Design}
\label{sec:design}


We first formulate token routing for minimizing the number of activated experts as an optimization problem (\S\ref{ssec:problem_formulation}). Unfortunately, we find that while an optimal algorithm exists for the above problem, it incurs prohibitively high computational overhead in practice (\S\ref{ssec:exact_algorithm}). To overcome this limitation, we propose \name, a routing algorithm that achieves \textit{near-optimal} routing quality with \textit{minimal overhead} (\S\ref{ssec:metro_algorithm}).

\subsection{Problem Formulation}~\label{ssec:problem_formulation}

We formulate token routing (dubbed \rproblem) as an Integer Linear Program (ILP), as detailed next.

\paragraphb{\rproblem} We consider $N$ experts across $G$ GPUs, with the binary expert--GPU mapping $A \in \{0,1\}^{N \times G}$ identifying which GPU hosts each expert. We define the number of tokens per expert as $T[1..N]$ for a particular batch, where $T[i]$ corresponds to the number of tokens for expert $i$ in that batch.
The problem of minimizing the number of activated experts reduces to finding assignments for: 
\begin{itemize}
    \item integer variables $x_{i,g} \ge 0$, which denote tokens of expert $i$ routed to GPU $g$,
    \item binary variables $y_{i,g} \in \{0,1\}$, which identifies whether expert $i$ is activated on GPU $g$, and,
    \item the integer variable $\lambda \ge 0$, which denotes the maximum number of activated experts across all GPUs,
\end{itemize}  
to minimize $\lambda$:
\[
  \min \ \lambda
\]
subject to, for all $g \in [G]$ and $i \in [N]$,
\begin{align}
  \sum_{i=1}^{N} y_{i,g} &\le \lambda, \label{eq:ilpmin-perGroupToken}\\
  \sum_{g=1}^{G} x_{i,g} &= T[i], \label{eq:ilpmin-flowConservation}\\
  x_{i,g} = y_{i,g} = 0 &\quad \text{if } A_{i,g} = 0, \label{eq:ilpmin-allowed}\\
  x_{i,g} &\le T[i] \cdot y_{i,g}. \label{eq:ilpmin-link}
\end{align}

Intuitively, constraint~(\ref{eq:ilpmin-perGroupToken}) ensures that the number of activated experts for each GPU does not exceed the upper bound $\lambda$. Constraint~(\ref{eq:ilpmin-flowConservation}) ensures no tokens are dropped, \ie, all $T[i]$ tokens of expert $i$ must be routed. Constraint~(\ref{eq:ilpmin-allowed}) ensures that routing abides by the placement matrix $A$, \ie, tokens should not be routed to GPUs that do not host the corresponding expert. Finally, constraint~(\ref{eq:ilpmin-link}) ensures that tokens are routed to a GPU ($x_{i, g} > 0$) only if the corresponding expert is activated there ($y_{i, g}>0)$.

\paragraphb{Simplifying \rproblem} Interestingly, we find that all feasible solutions to \rproblem share a common property that allows us to focus on a simpler variant of \rproblem:

\begin{lemma}
\label{lemma:one_replica_per_expert}
Any feasible solution to \rproblem either routes tokens only to one replica of any expert, or can be mapped to a solution that does without increasing the objective value.
\end{lemma}

\emph{Proof sketch.}
Consider any feasible solution that achieves an objective value of $\lambda=\lambda_0$. In this solution, if the tokens to be processed by any expert $i$ are distributed across multiple GPUs (\ie, its replicas), we construct another solution in which all such experts $i$'s tokens are routed to a single GPU (\ie, a single replica). This solution preserves constraints~(\ref{eq:ilpmin-flowConservation}), (\ref{eq:ilpmin-allowed}), and (\ref{eq:ilpmin-link}). At the same time, it does not increase the number of activated experts on any GPU, \ie, the solution still preserves constraint~(\ref{eq:ilpmin-perGroupToken}) and thus still achieves $\lambda\leq\lambda_0$, proving the second half of the lemma's claim. If all tokens in the original feasible solution are already routed to a single replica of each expert, then the first part of the lemma's claim holds, completing our proof.\qed

\vspace{0.1in}
\noindent
Leveraging Lemma~\ref{lemma:one_replica_per_expert}, we restrict our focus to solving the simpler problem of identifying for each expert (that has tokens to process) one of its replicas to activate, such that the maximum number of activated expert replicas per GPU is minimized. This is because any feasible solution to \rproblem can be reduced to a feasible solution to this simpler problem. We subsequently use \rproblem to refer to this simpler problem variant. 

\begin{figure}[t!]
  \centering
 \includegraphics[width=0.28\textwidth]{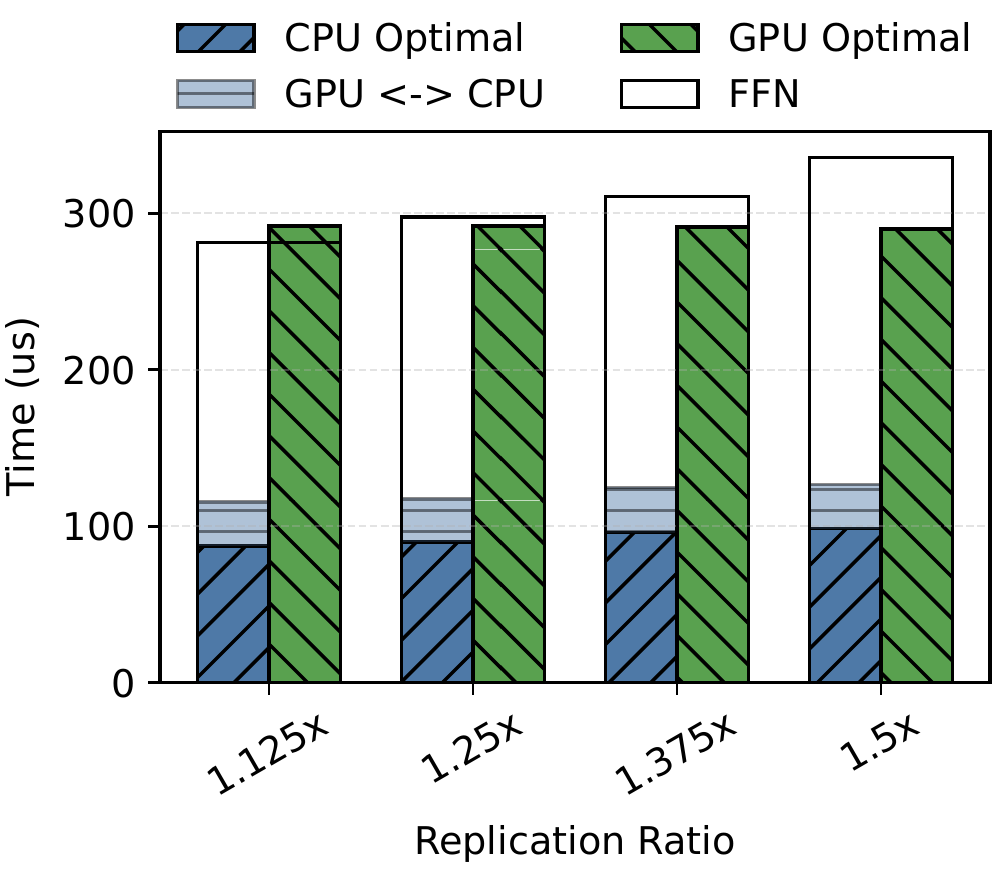}
  \caption{\textbf{The runtime of CPU-based and GPU-based optimal algorithms VS. the runtime of a single FFN layer of Qwen3-30B on vLLM over 8 A100 GPUs.} The CPU-based and GPU-based algorithms incur prohibitively high overheads of $31.4\%$ – $41.3\%$ and $86.4\%$ – $103.8\%$ relative to the FFN, respectively.}\label{fig:design_optimal_time}
\end{figure}

\subsection{An Optimal Algorithm}
\label{ssec:exact_algorithm}

We find that \rproblem reduces to the classical optimization problem~\cite{exact_algo_1, exact_algo_2, exact_algo_3} of matching jobs of identical runtime (\ie, experts) to machines (\ie, GPUs) with restriction (\ie, expert-GPU placement matrix $A$) to minimize the maximum runtime across machines (\ie, maximum activated experts across GPUs). Similar to prior approaches for solving such problems~\cite{exact_algo_1, exact_algo_2, exact_algo_3}, our optimal algorithm searches for the minimal feasible objective (\ie, $\lambda$) via binary search, where each candidate $\lambda$ is tested for feasibility via bipartite matching~\cite{bipartite_matching}. Specifically, for a particular candidate $\lambda_0$, we can build a bipartite graph with experts (that process at least one token in this batch) on the left and GPUs on the right. Edge $(i,g)$ is present iff expert $i$ is placed on GPU $g$ (\ie, $A_{i,g}=1$). Unlike the standard bipartite matching, where each node is matched at most once, our setting requires a variant in which each GPU $g$ node can be matched at most $\lambda_0$ times. This variation accounts for the fact that each GPU can activate at most $\lambda_0$ expert replicas. With this reduction, a feasible bipartite matching in this graph exists iff $\lambda_0$ is feasible for \rproblem.

\paragraphb{Computational overheads} From a computational complexity perspective, the binary search incurs a complexity factor of $O(\log{\lceil \frac{|A|}{G} \rceil})$ since $\lambda$ is at most $\lceil \frac{|A|}{G} \rceil$. For each candidate $\lambda$, a bipartite-matching-based feasibility test using max-flow algorithms incurs a worst-case complexity of $O((N + G)^2\cdot(\lceil \frac{|A|}{G} \rceil + N + G))$. 

To study this overhead in practice, we measure the runtime of both the state-of-the-art CPU-based and GPU-based implementations of the optimal algorithm. In our implementations, we use Dinic's max-flow algorithm~\cite{dinic1970algorithm} for feasibility testing on the CPU, and a recent, state-of-the-art, GPU-optimized push-relabel max-flow algorithm~\cite{vanausdalefficient} on the GPU. Unfortunately, we find that the optimal algorithm, whether executed on CPU or GPU, has prohibitively high computational overhead relative to the model's computation time. Fig.~\ref{fig:design_optimal_time} shows that the CPU-based and GPU-based optimal algorithms incur runtime overheads of $116.3\mu s$ – $128.8\mu s$ and $290.0\mu s$ – $292.1 \mu s$, respectively, across different replication ratios. Relative to the runtime of a single FFN layer of Qwen3-30B on vLLM over 8 A100 GPUs, these correspond to prohibitively high overheads of $31.4\%$ – $41.3\%$ and $86.4\%$ – $103.8\%$, respectively. The GPU-based algorithm is slower than the CPU-based one because the bipartite graph is insufficiently large to fully exploit the GPU’s parallelism. Furthermore, we exclude any CPU-based algorithms because transferring the algorithm inputs (\ie, the top-$k$ selection tensors, which reside on the GPU) alone costs $26.5 \mu s$ – $29.2 \mu s$, introducing up to $10.4\%$ overhead relative to the FFN.

\begin{algorithm}[t]
\caption{\name's greedy approximate algorithm}\label{alg:greedy_parallel_2pl}
\begin{algorithmic}[1]
\STATE \textbf{Input:} $N$, $G$, $A \in \{0,1\}^{N \times G}$, $T[1..N]$
\STATE \textbf{Output:} $y_{i,g}$
\STATE \textbf{Initialization:} $L[g] \gets 0$ and init lock $l_g$ for each $g = 1,\dots,G$; $y_{i,g} \leftarrow 0$ for each $g = 1,\dots,G$, $i = 1,\dots,N$.\label{line:greedy_init}
\STATE \textbf{for} $i = 1$ to $N$ \textbf{do in parallel}
    \STATE \quad \textbf{if} $T[i] > 0$
    \STATE \quad\quad $\mathcal{G}_i \gets \{ g \mid A_{i,g} = 1 \}$\label{line:greedy_gpuset}
    \STATE \quad\quad acquire all locks $\{ \ell_g \mid g \in \mathcal{G}_i \}$ in a total order\label{line:greedy_lock}
    \STATE \quad\quad choose $g^\star \in \mathcal{G}_i$ with  the smallest $L[g]$\label{line:greedy_choose}
    \STATE \quad\quad $y_{i,g^\star} \gets 1$; \quad $L[g^\star] \gets L[g^\star] + 1$\label{line:greedy_assign}
    \STATE \quad\quad release all locks $\{ \ell_g \mid g \in \mathcal{G}_i \}$\label{line:greedy_unlock}
    \STATE \quad \textbf{end if}
\STATE \textbf{end for}
\end{algorithmic}
\end{algorithm}

\begin{figure}[t!]
  \centering
 \includegraphics[width=0.475\textwidth]{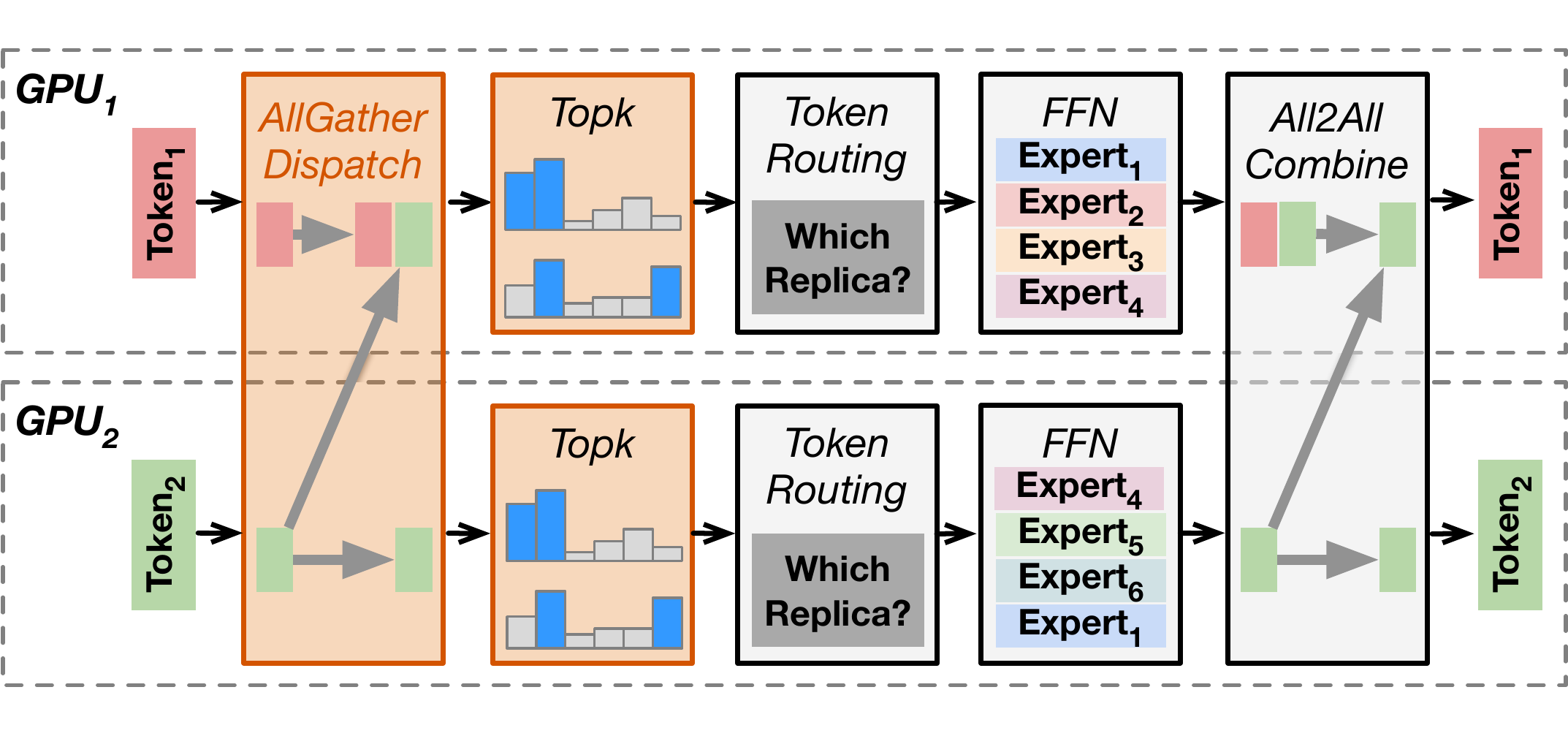}
  \hfill
  \caption{\textbf{\name replaces the conventional \alltoall with \allgather dispatch before top-$k$ for every GPU to obtain the global top-$k$ knowledge as input to Algorithm~\ref{alg:greedy_parallel_2pl}, with minimal overhead~(\ref{ssec:metro_algorithm}).}}\label{fig:design_allgather}
\end{figure}

\subsection{\name: A Greedy Approximate Algorithm}\label{ssec:metro_algorithm}



To simultaneously achieve low computational overhead while preserving near-optimal routing quality, we propose \name, a GPU-native approximate algorithm that prioritizes computational efficiency. In addition, \name introduces a novel \allgather communication scheme that collects the global top-$k$ required by the algorithm while incurring insignificant overhead in practice.

\paragraphb{\name's greedy approximate algorithm} Algorithm~\ref{alg:greedy_parallel_2pl} details \name's greedy approximate algorithm. It inputs a problem instance of \rproblem, and outputs the assignments for decision variables $y_{i,g}$ that denote whether expert $i$ is activated on GPU $g$ in the routing decision. We omit other decision variables (\ie, $x_{i,g}$ and $\lambda$) since Lemma~\ref{lemma:one_replica_per_expert} allows us to directly compute them using $y_{i,g}$: 
$$
x_{i,g} = 
\begin{cases} 
T[i] & \text{if} y_{i,g} = 1\\
0 & \text{otherwise}
\end{cases}
$$
$$\lambda = \max_{g=1 \dots G} \sum_{i=1}^{N} y_{i,g}$$

Algorithm~\ref{alg:greedy_parallel_2pl} maintains a per-GPU activated expert counter $L[1..N]$ that is guarded by corresponding locks $l_{1..N}$~(line~\ref{line:greedy_init}). The high-level idea is to greedily assign each expert to the GPU with the fewest activated experts.
In more detail, for each expert that has at least one token to process for this batch, \name first identifies its candidate GPUs by looking up the expert-GPU placement matrix $A$~(line~\ref{line:greedy_gpuset}). Then, before matching experts to GPUs, to ensure the atomicity of each expert's assigning process, each thread acquires locks for its expert's all candidate GPUs~(line~\ref{line:greedy_lock}), and only releases them after the expert assignment is done~(line~\ref{line:greedy_unlock}). To avoid deadlocks, lock acquisition is done in a total order (e.g., GPU ID order in our implementation). After all locks are acquired, each expert chooses the GPU $g$ that has the least activated experts $L[g]$~(line~\ref{line:greedy_choose}) and assigns itself to $g$~(line~\ref{line:greedy_assign}). We defer architecture-specific implementation details to \S\ref{sec:impl}.

\paragraphb{All-gathering Global Top-$k$ Knowledge}
\name requires global top-$k$ information ($T[1..N]$) across all GPUs, which conflicts with the conventional \alltoall dispatch where top-$k$ results remain local to each GPU. To address this, as shown in Fig.~\ref{fig:design_allgather}, we replace \alltoall with an \allgather dispatch before top-$k$ so that every GPU can obtain the global top-$k$ knowledge with minimal overhead. Specifically, as shown in Fig.~\ref{fig:design_allgather}, tokens are first all-gathered to all GPUs; each GPU then runs top-$k$ on the full token set to build the global top-$k$ knowledge (\ie, $T[1..N]$). Then, each GPU executes Algorithm~\ref{alg:greedy_parallel_2pl} to route tokens, compute FFNs, and finally performs the \alltoall combine.

\paragraphb{Minimal performance overhead} \name introduces three sources of performance overheads, all of which are outweighed by the performance improvement it enables by effectively minimizing experts. While we break down these overheads empirically in \S\ref{ssec:endtoend_perf}, we provide an intuitive analysis of \name's overheads here. First, Algorithm~\ref{alg:greedy_parallel_2pl} runs in $O(|A|)$ time, since the placement of each export on a GPU is considered exactly once. This is substantially lower than the optimal algorithm’s $O\big((N+G)^2 \cdot (\lceil |A|/G \rceil + N + G) \cdot \log \lceil |A|/G \rceil\big)$. Second, while the \allgather dispatch requires redundant top-$k$ over the full token set on each GPU, top-$k$ computation time is negligible compared to other components (\ie, attention, FFN, and communication). Consequently, our empirical results show that the overheads introduced by redundant top-$k$ computations in \name remain negligible. Lastly, while \name's \allgather dispatch is theoretically more expensive than the conventional \alltoall, we observe no statistically significant increase in communication time. This is because in the memory-bound regime (\ie, small batches), NVLink latency dominates while the bandwidth cost is minor: with 32 decode tokens per GPU (8 GPUs) and the $fp16$ data type, \alltoall sends $256$KB/GPU and \allgather sends $2$MB/GPU, which on 600 GB/s NVLink translates to only $\sim400ns$ and $\sim3 \mu s$, respectively --- far below the tens to $\sim100 \mu s$ fixed cost of launching NCCL collectives.

\begin{figure}[t!]
  \centering
 \includegraphics[width=0.5\textwidth]{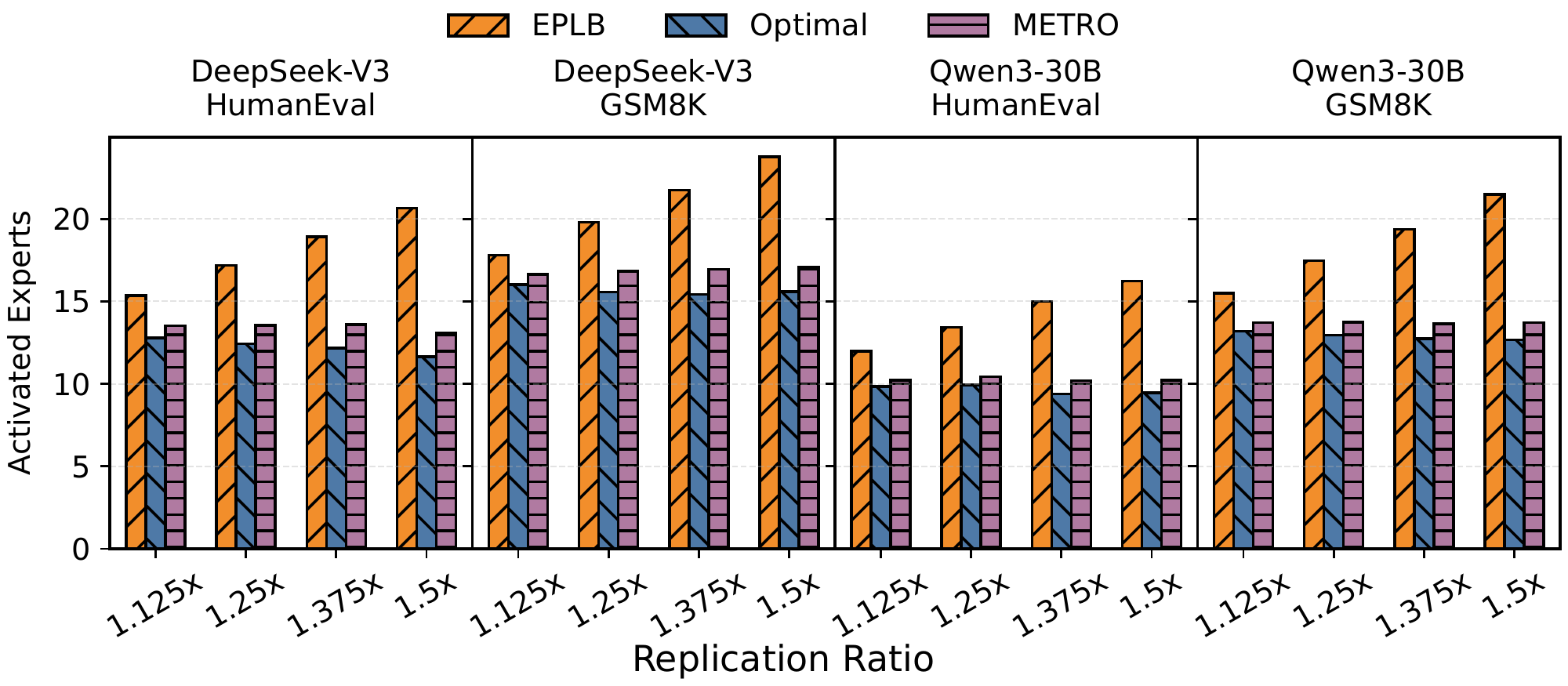}
  \caption{\textbf{The maximum number of activated experts per GPU per decode batch ($32$ tokens) for EPLB routing, the optimal algorithms, and \name across different models (\ie, DeepSeek-V3 and Qwen3-30B) and datasets (\ie, Humaneval and GSM8K) (\S\ref{ssec:exact_algorithm}).} \name is within $10.9\%$ higher than the optimal algorithm (\ie, optimal routing) and is lower than EPLB by up to $42.3\%$.}\label{fig:design_optimal_vs_greedy_experts}
\end{figure}

\paragraphb{Near-optimal routing quality} \name's greedy algorithm achieves near-optimal routing quality. 
Fig.~\ref{fig:design_optimal_vs_greedy_experts} shows that \name's maximum number of activated experts per GPU per decode batch ($32$ tokens) is within $10.9\%$ of the optimal algorithm across different models (\ie, DeepSeek-V3 and Qwen3-30B), different datasets (\ie, Humaneval and GSM8K), and replication ratios. This is up to $42.3\%$ lower than EPLB's routing algorithm that evenly distributes tokens across replicas.












%% file: tex/implementation.tex
\section{Implementation}
\label{sec:impl}

\paragraphb{CUDA Graph integration}
To integrate \name into vLLM with no additional GPU kernel launch overheads, we take advantage of vLLM’s compilation framework~\cite{vllm-cuda-graphs} to integrate \name into its decode phase CUDA Graphs. We precompile graphs for power-of-two batch sizes --- up to $32$ tokens per GPU --- which cover all batch sizes considered in our evaluations. For non-power-of-two batches, vLLM pads to the next power-of-two and reuses the corresponding graph.

\paragraphb{Efficient GPU kernel implementation}
We implement \name (\ie, Algorithm~\ref{alg:greedy_parallel_2pl}) as a CUDA kernel that runs on a single streaming multiprocessor (SM). Since the algorithm’s parallelism is fundamentally bounded by the number of experts (e.g., $128$ for Qwen and $256$ for DeepSeek-V3) and locking further reduces concurrency to below $64$, a single A100 SM provides sufficient parallel processing power. Confining the kernel to a single SM also lets us keep the load counters $L$ and locks $l$ in SM-local shared memory for faster access. We use a simple test-and-set lock~\cite{taslock} for synchronization.

%% file: tex/evaluation.tex
\section{Evaluation}
\label{sec:eval}

We evaluate \name's performance improvements relative to the state-of-the-art EP load balancer atop both a real system implementation and a proprietary industrial performance simulator. Our simulation studies allow us to effectively explore the vast design space of different models, datasets, and hardware configurations, without being limited by the cost- and time-inefficiencies of a real-system implementation.





\subsection{Setup}
\label{ssec:system_setup}

\begin{table}[t!]
\centering
\caption{Simulation Setup}
\label{tab:dlsim_config}
\begin{tabular}{lcc}
\hline
\textbf{Parameter} & \textbf{Specification} \\

\hline
\hline
 & \textbf{Real System} \\
\hline
Models & Qwen3-30B-A3B \\
Datasets & NuminaMath (Decode-heavy),\\ &InstructCoder (Decode-heavy) \\
GPU Arch & 8 NVIDIA A100 40GB GPUs \\
NVLink & 600 GB/s, all GPUs within the same domain \\
\hline
\hline
 & \textbf{Simulation} \\
\hline
Models & Qwen3-235B-A22B, Deepseek-V3-671B \\

Datasets & GSM8K (Prefill-heavy), Humaneval (Decode-heavy) \\
Modeled GPU Arch & 8/16 NVIDIA B200 192GB GPUs\\
NVLink & 900 GB/s, all GPUs within the same domain \\
\hline
\end{tabular}
\end{table}

Our experimental setup is summarized in Table~\ref{tab:dlsim_config}.


\paragraphb{Models} 
For our real-system evaluation, we use Qwen3-30B-A3B~\cite{qwen3technicalreport} as a representative fine-grained MoE (\ie, $128$ experts) that fits the memory capacity of our GPU node. 

For simulation studies, we used two models: Qwen3-235B-A22B~\cite{qwen3technicalreport} ($128$ experts), and Deepseek-V3-671B~\cite{deepseekai2025deepseekv3technicalreport} ($256$ experts). 
We expect our simulation studies to validate the trends and insights from the real system implementation with larger models.

\paragraphb{Workloads and traces} 
We evaluate our real-system implementation using two workloads: InstructCoder~\cite{li2024instructcoderinstructiontuninglarge} and NuminaMath~\cite{numina_math_datasets}. InstructCoder is a code-editing instruction dataset of over $114K$ instruction + input-code + output-code triplets used to assess code generation models. NuminaMath is a large-scale dataset (\ie, ~$900K$) of competition-level math problems with chain-of-thought reasoning, spanning high-school contests, Olympiads, and international sources, for post-training reasoning models. 

Our simulation studies use Humaneval~\cite{chen2021evaluating} and GSM8K~\cite{cobbe2021gsm8k}. Humaneval is a benchmark of $164$ Python programming problems used to assess code generation models. GSM8K is a collection of $8,500$ grade-school-level math word problems for evaluating the mathematical reasoning of language models. 

We chose NuminaMath, Humaneval, and InstructCoder as representative decode-heavy workloads, and GSM8K as a prefill-heavy workload, to demonstrate \name's benefits across different workload types.

\begin{figure*}[t!]
  \centering
    \subfloat[InstructCoder]{\includegraphics[width=0.5\textwidth]{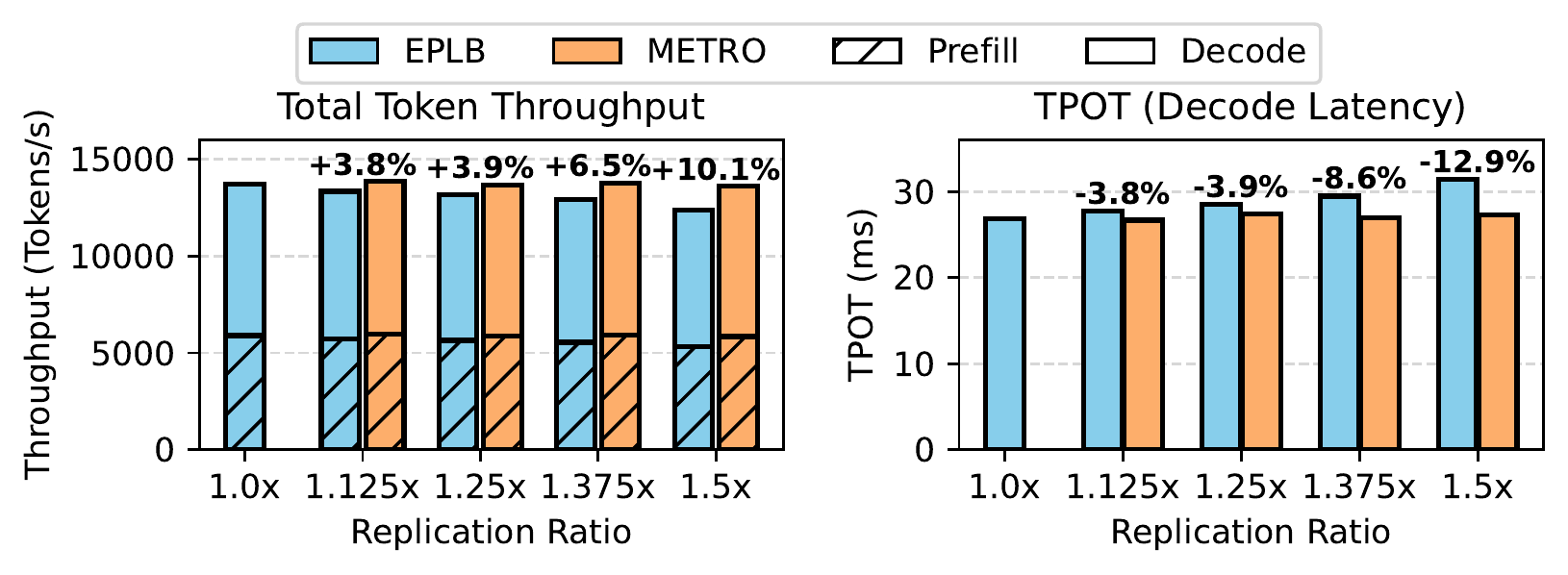}\label{fig:decode_and_throughput_instructcoder_summary_2bar}}
    \hfill
    \subfloat[NuminaMath]{\includegraphics[width=0.5\textwidth]{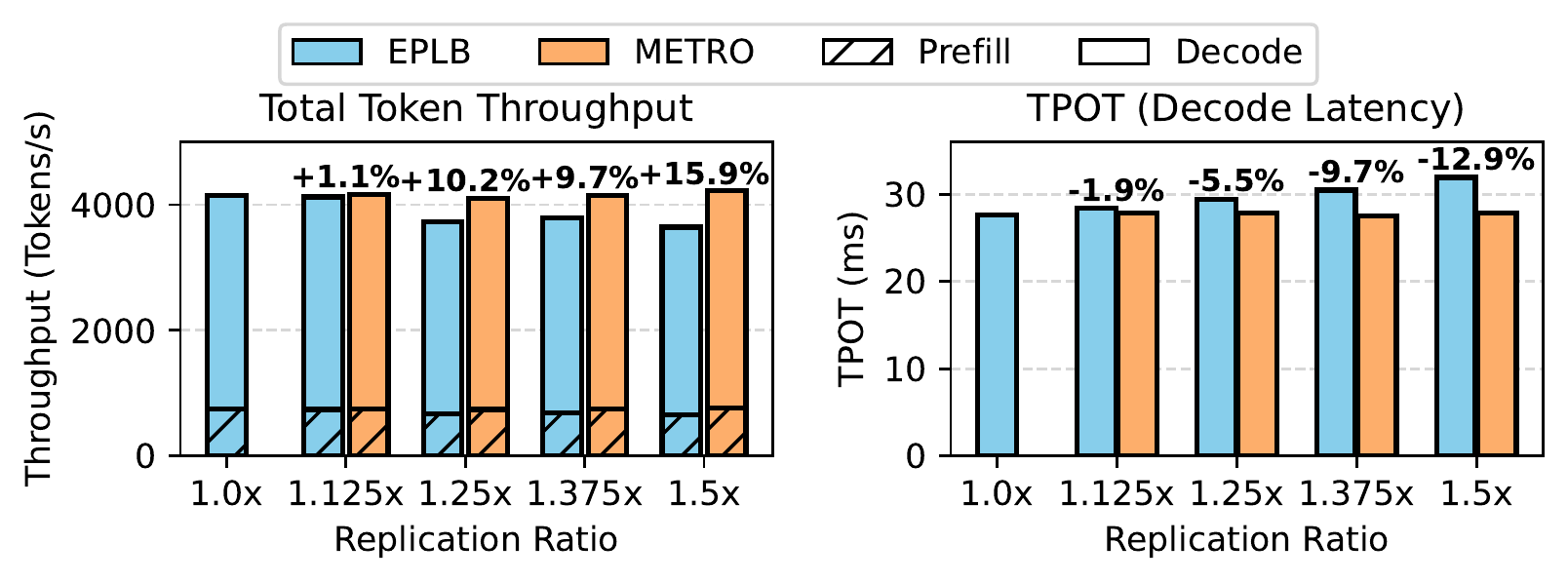}\label{fig:decode_and_throughput_numinamath_summary_2bar}}
    \hfill
    
  \caption{\textbf{Total token throughput and decode latency (\ie, Time-per-Output-Token, TPOT) for Qwen3-30B and DeepSeek-V3 across InstructCoder and NuminaMath datasets (\S\ref{ssec:system_setup}), with full EP and decode batch size = $32$ per GPU.} \name improves throughput by up to $15.9\%$, and reduces decode latency by up to $12.9\%$ at $1.5\times$ replication, compared to EPLB.}
  \label{fig:throughput_latency_comparison_realsys}
\end{figure*}

\begin{figure*}[t!]
  \centering
    \subfloat[Qwen3-235B, Humaneval, 8*B200]{\includegraphics[width=0.5\textwidth]{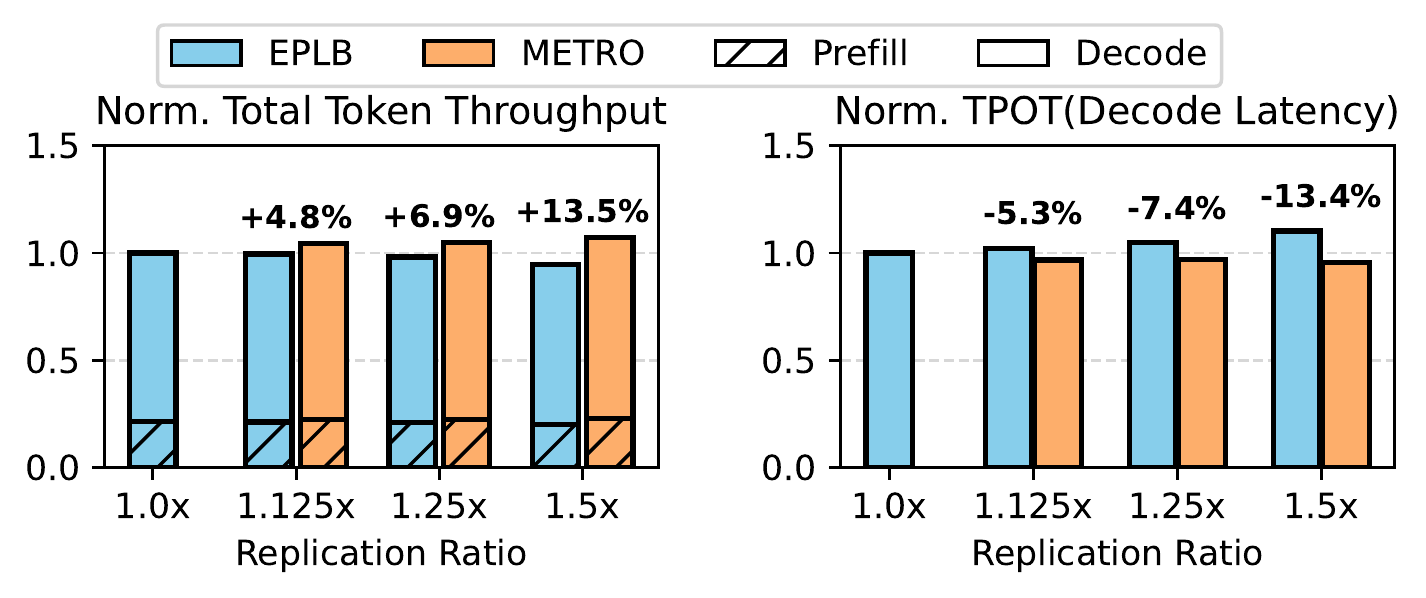}\label{fig:throughput_latency_comparison_qwen_humaneval}}
    \hfill
    \subfloat[Qwen3-235B, GSM8K, 8*B200]{\includegraphics[width=0.5\textwidth]{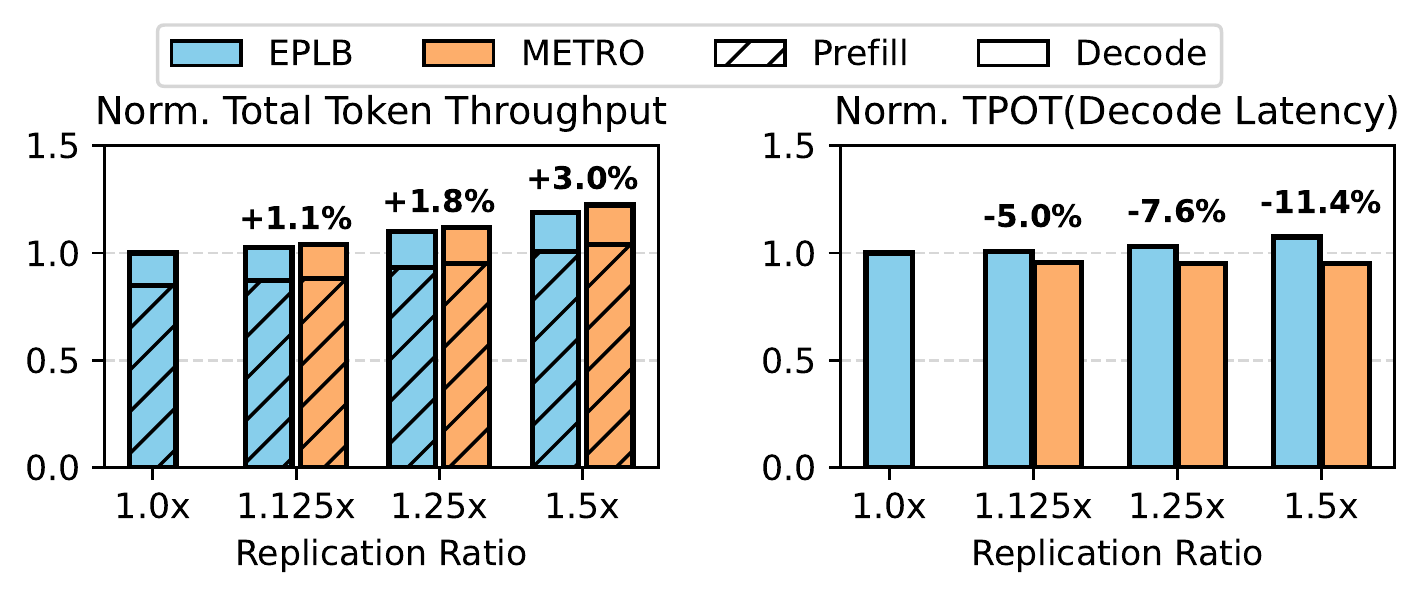}\label{fig:throughput_latency_comparison_qwen_gsm8k}}
    \hfill
    \subfloat[DeepSeek-V3, Humaneval, 16*B200]{\includegraphics[width=0.5\textwidth]{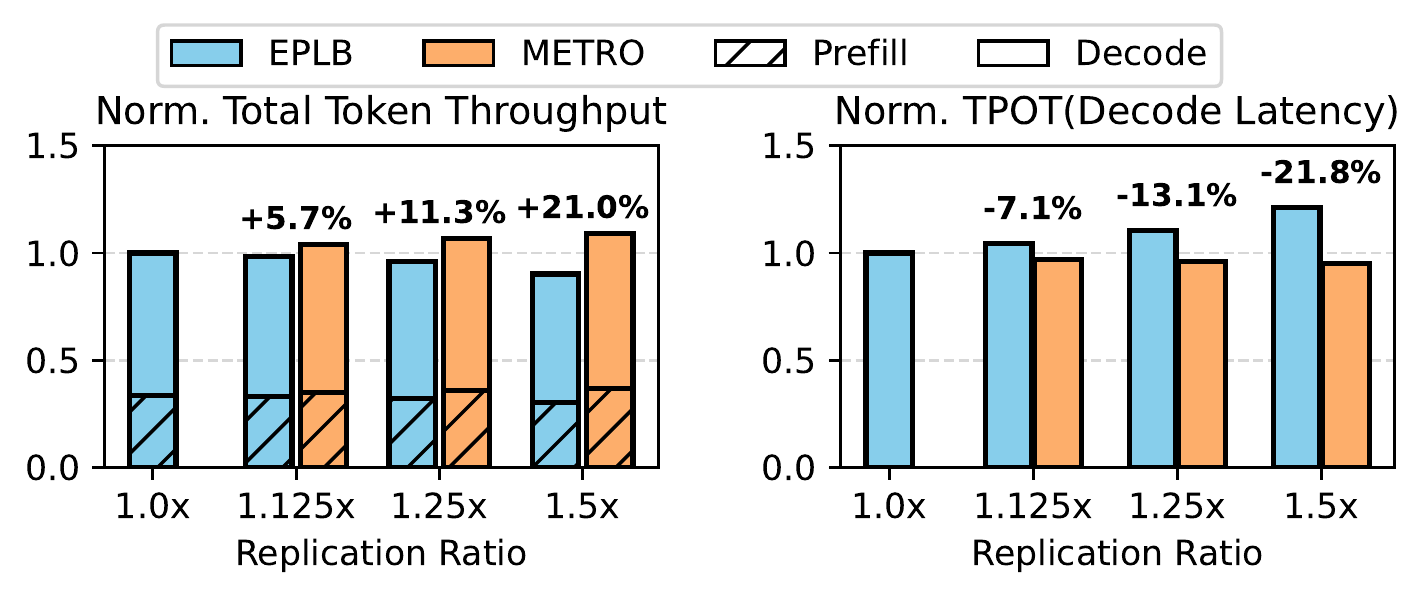}\label{fig:throughput_latency_comparison_deekseek_humaneval}}
    \hfill
    \subfloat[DeepSeek-V3, GSM8K, 16*B200]{\includegraphics[width=0.5\textwidth]{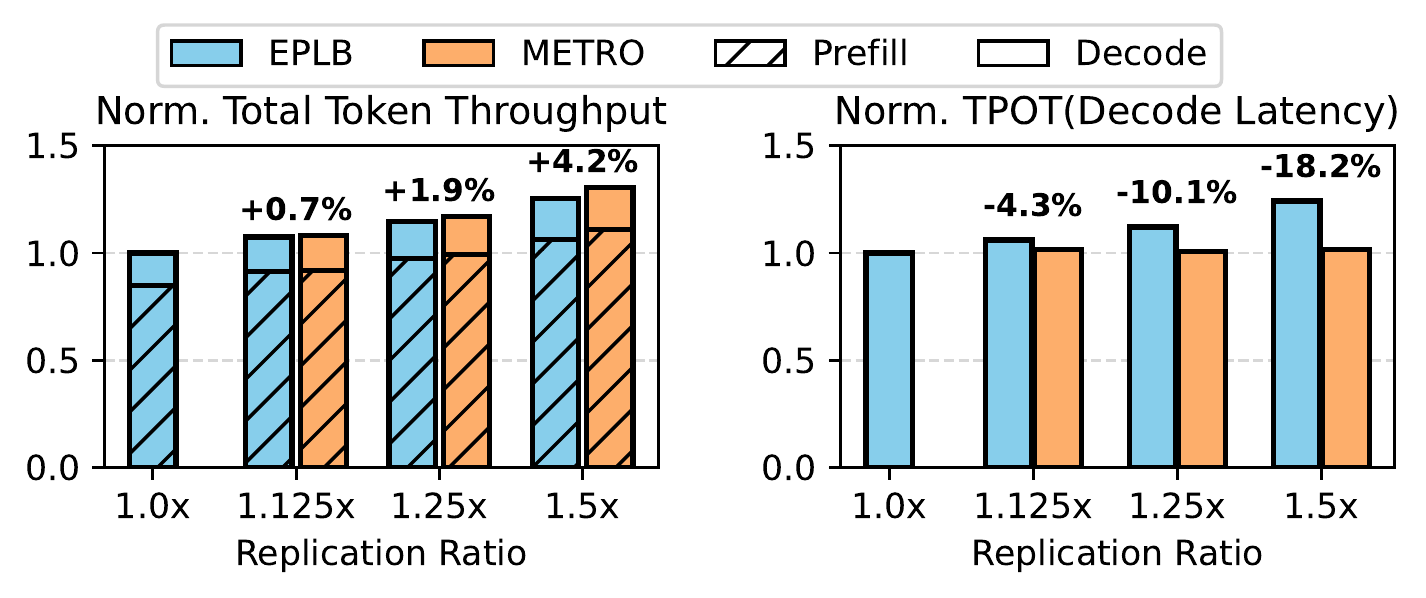}\label{fig:throughput_latency_comparison_deepseek_gsm8k}}
    \hfill
    
  \caption{\textbf{Simulated total token throughput and decode latency (Time-per-Output-Token, TPOT) for Qwen3-235B and DeepSeek-V3 models across GSM8K and Humaneval datasets (\S\ref{ssec:system_setup})}. Y-axis is normalized to EPLB with $1.0x$ memory capacity due to confidentiality. Full EP and global decode batch size = $1024$. With EPLB routing, more replication improves the prefill-heavy workload (GSM8K)'s throughput but hurts the decode-heavy workload (Humaneval)'s throughput. \name's consistent benefit on decode latency (up to $21.8\%$ at $1.5\times$ replication) improves total token throughput to outperform EPLB routing for every replication ratio (up to $21.0\%$ at $1.5\times$ replication).}
  \label{fig:throughput_latency_comparison}
\end{figure*}

\paragraphb{Hardware and software setup}
Our real-system implementation employed vLLM on a Google Cloud a2-highgpu-8g VM instance equipped with 8 NVIDIA A100 40GB GPUs interconnected by 600GB/s NVLink. We used data parallelism for the attention layers and EP for the expert FFN. We limited the maximum batch size for the decode phase to $32$ tokens per GPU and the maximum prompts for the prefill phase to $32$ per GPU. We set the context length to $8K$.

Our simulation studies modeled NVIDIA B200 GPUs~\cite{nvidia_blackwell_tech_overview_2025}, the state-of-the-art GPU production-optimized for LLM workloads. To accommodate full model parameters while maintaining efficient GPU utilization, we used eight B200 GPUs (as configured in the NVIDIA DGX B200 Systems~\cite{nvidia_dgx_b200_web}) for the Qwen experiments and sixteen B200 GPUs for the DeepSeek experiments. In both setups, all GPUs are interconnected by 900GB/s NVLink. We set the sequence length to $1K$ (input) and $2K$ (output) for the decode phase. 
We set the global decode batch size to $1K$, and simulate chunked prefill~\cite{agrawal2024sarathi_serve} to limit prefill batch size to $8K$ to meet a reasonable service-level objective (SLO). 
We used full EP across all GPUs, which yields the best performance for both prefill and decode at the chosen batch size compared to tensor parallelism (TP). Evaluation across a broader range of batch sizes and parallelism settings (\S\ref{ssec:decode_only_analysis}) shows that our insights from the default parameters generalize to a wide range of configurations. 

\paragraphb{Compared baselines}
We compare \name against EPLB's routing algorithm that evenly distributes tokens across expert replicas. We do not modify the expert placement and replication scheme to avoid any interference with the latency of the prefill phase --- both \name and EPLB routing algorithms use EPLB's expert placement and replication. $1.0\times$ replication represents EPLB placement with no replication (\ie, no token-routing strategy is needed). Note that \name is only applied to the decode (\ie memory-bound) phase and the prefill phase still uses EPLB's token routing.

\paragraphb{Simulator description} We used a proprietary industrial simulator, which is a fine-grained analytical roofline model designed to capture detailed performance behaviors of large-scale GPU systems. 
The simulator supports multi-GPU configurations and accounts for workload imbalance by estimating the runtime based on the performance of the most bottlenecked GPU. It models hardware architecture across multiple levels, including register, shared memory, compute, L2, HBM, and network operations. Its network component computes data transfer volumes to estimate throughput, latency, and potential congestion effects.
In addition, it models various mapping strategies across different parallelism levels, including tensor parallelism (TP) and expert parallelism (EP). While the overall silicon validation has been conducted internally to ensure modeling accuracy and reliability, we are unable to share its details due to NDA constraints.

\paragraphb{Simulation traces} 
For each workload, we collected real traces from running the model with token routing information for each layer. We used the replayed trace to identify two key metrics: the maximum number of tokens processed per GPU and the maximum number of activated experts per GPU. 
We fed these as the inputs to the simulator to describe the workload on the bottleneck GPU and derive the runtime.

\subsection{End-to-end Performance Analysis}
\label{ssec:endtoend_perf}

\paragraphb{Overall performance} 
Fig.~\ref{fig:throughput_latency_comparison_realsys} and Fig.~\ref{fig:throughput_latency_comparison} show our results for the real system and simulation, respectively. We consider two metrics: Total Token Throughput for prefill and decode when they are co-deployed, and Decode Latency (Time-per-Output-Token, TPOT).
Compared to EPLB, \name reduces decode latency by $1.9\%$ - $21.8\%$, and improves throughput by $0.7\%$ - $21.0\%$, and , across replication ratio $1.125\times$ - $1.5\times$ and datasets.
Moreover, \name's performance improvements increase with more replication due to EPLB's increased number of activated experts, which degrades performance (increasing decode latency by as much as $20\%$ at $1.5\times$ replication compared to no replication).




\begin{tcolorbox}
\emph{Takeaway}: \name improves upon EPLB routing in both decode latency ($1.9\%$--$21.8\%$) and total token throughput ($0.7\%$--$21.0\%$) across different workloads, models, and hardware configurations. 
\end{tcolorbox} 

\paragraphb{Throughput implications for prefill- vs. decode-heavy workloads} 
Since \name primarily benefits the memory-bound decode phase, its performance implications for total token throughput can vary across workloads with different prefill-to-decode ratios. \name improves throughput by up to $21.0\%$ for decode-heavy workloads (Fig.~\ref{fig:decode_and_throughput_instructcoder_summary_2bar},~\ref{fig:decode_and_throughput_numinamath_summary_2bar},~\ref{fig:throughput_latency_comparison_qwen_humaneval},~\ref{fig:throughput_latency_comparison_deekseek_humaneval}), and up to $4.2\%$ for prefill-heavy workloads (Fig.~\ref{fig:throughput_latency_comparison_qwen_gsm8k},~\ref{fig:throughput_latency_comparison_deepseek_gsm8k}).

Since prefill is compute-bound, increasing the replication ratio with available memory capacity improves token balancing and prefill performance. For decode-heavy workloads, however, EPLB routing during the memory-bound decode phase negatively affects latency, leading to a notable drop in overall throughput compared to no replication. \name enables prefill to benefit from high replication while simultaneously improving decode performance, thereby significantly boosting overall throughput compared to EPLB routing. 

For prefill-heavy workloads, high replication yields better overall throughput with EPLB routing despite the performance degradation in decode. While decode does not consume the majority of runtime, applying \name to reduce decode latency still moderately improves throughput over EPLB routing.


\begin{figure}[t!]
  \centering
 \includegraphics[width=0.5\textwidth]{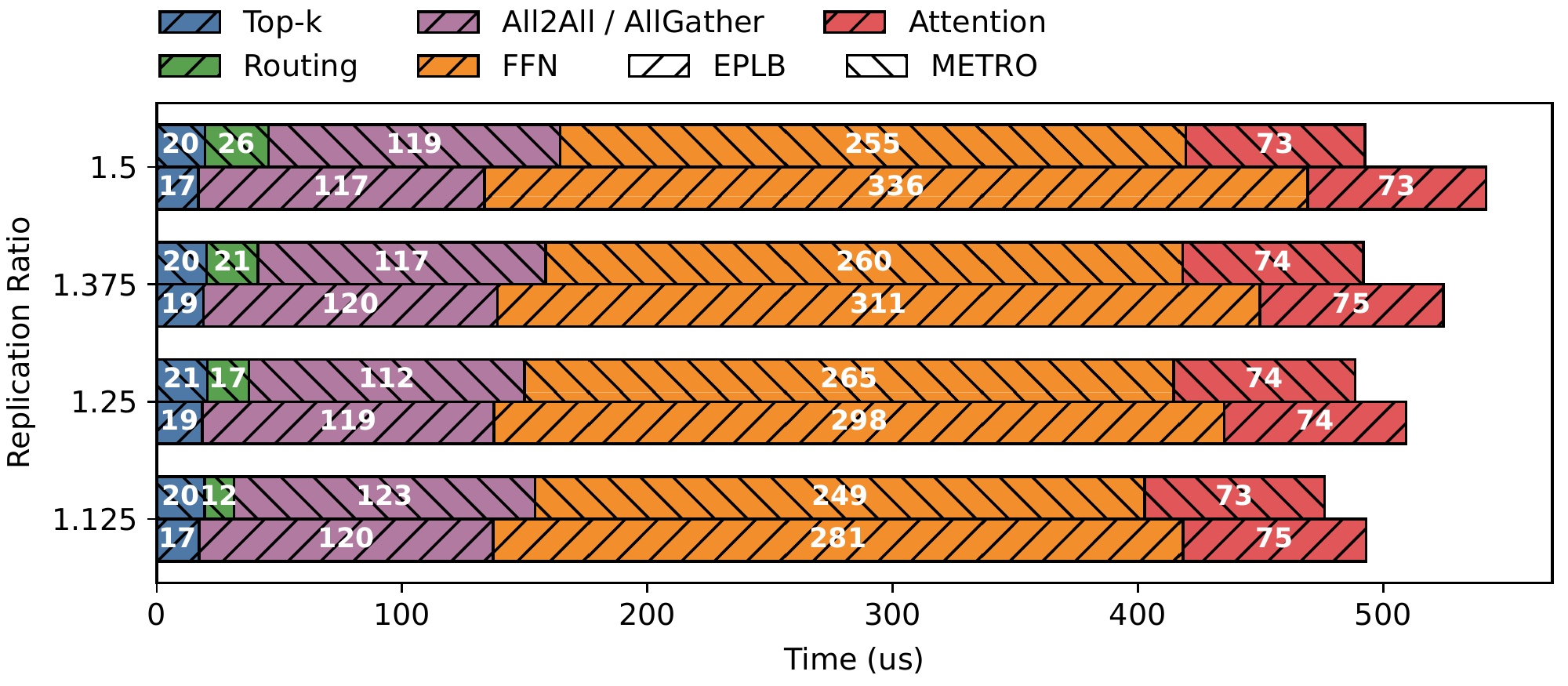}
  \caption{\textbf{Runtime breakdown for one layer of Qwen3-30B with various replication ratios (\S\ref{ssec:metro_algorithm}).} Results are averaged across all layers. \name's \allgather scheme introduces negligible overheads on top-$k$ and inter-GPU communication. \name's greedy approximate algorithm (\ie, Algorithm~\ref{ssec:metro_algorithm}) introduces minimal computational overhead (\ie, up to $26 \mu s$), which is more than offset by its FFN time reduction (\ie, up to $81 \mu s$). Consequently, \name reduces end-to-end decode latency by up to $\sim10\%$ compared to EPLB.}\label{fig:design_latency_breakdown}
\end{figure}

\begin{figure*}[t!]
  \centering
  \subfloat[Qwen3-235B, Humaneval, 8*B200]{\includegraphics[width=0.248\textwidth]{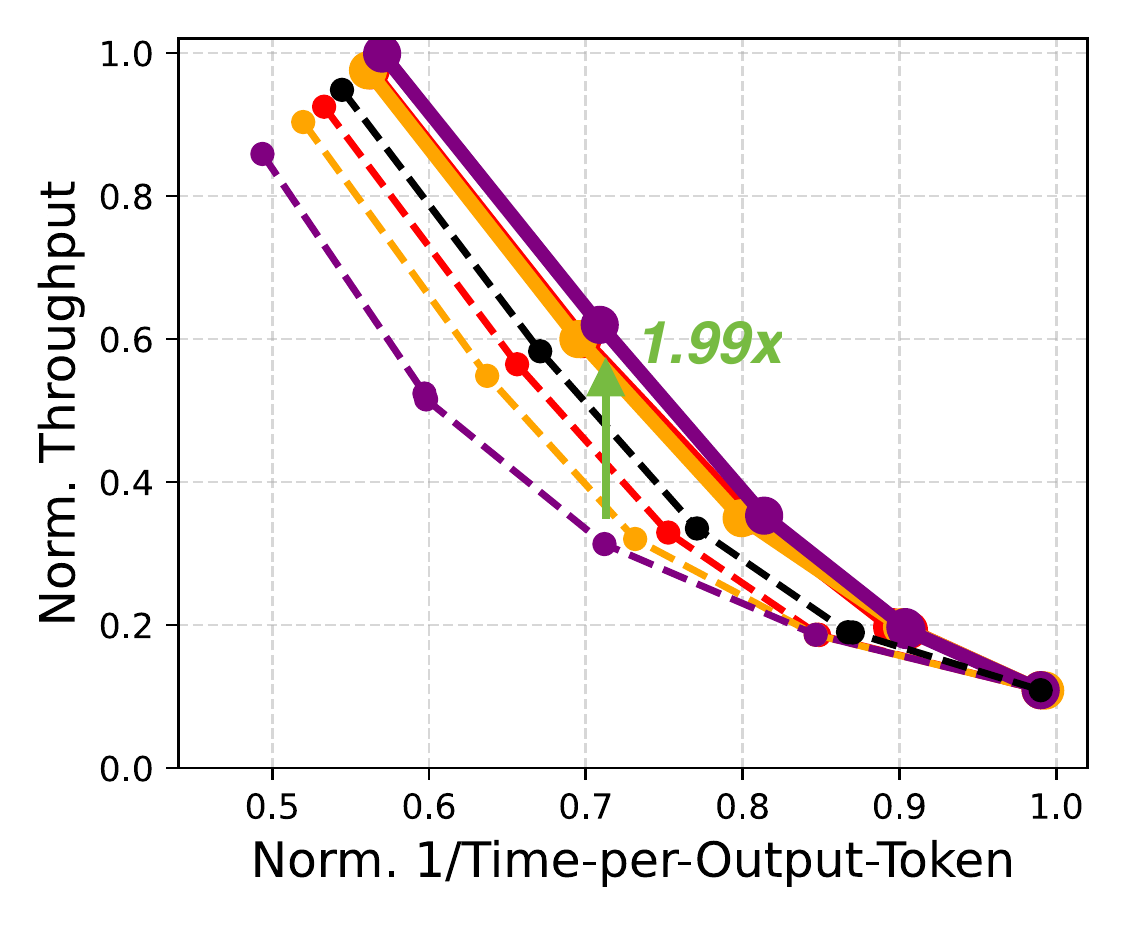}\label{fig:decode_pareto_curve_qwen_humaneval}}
  \hfill
    \subfloat[Qwen3-235B, GSM8K, 8*B200]{\includegraphics[width=0.248\textwidth]{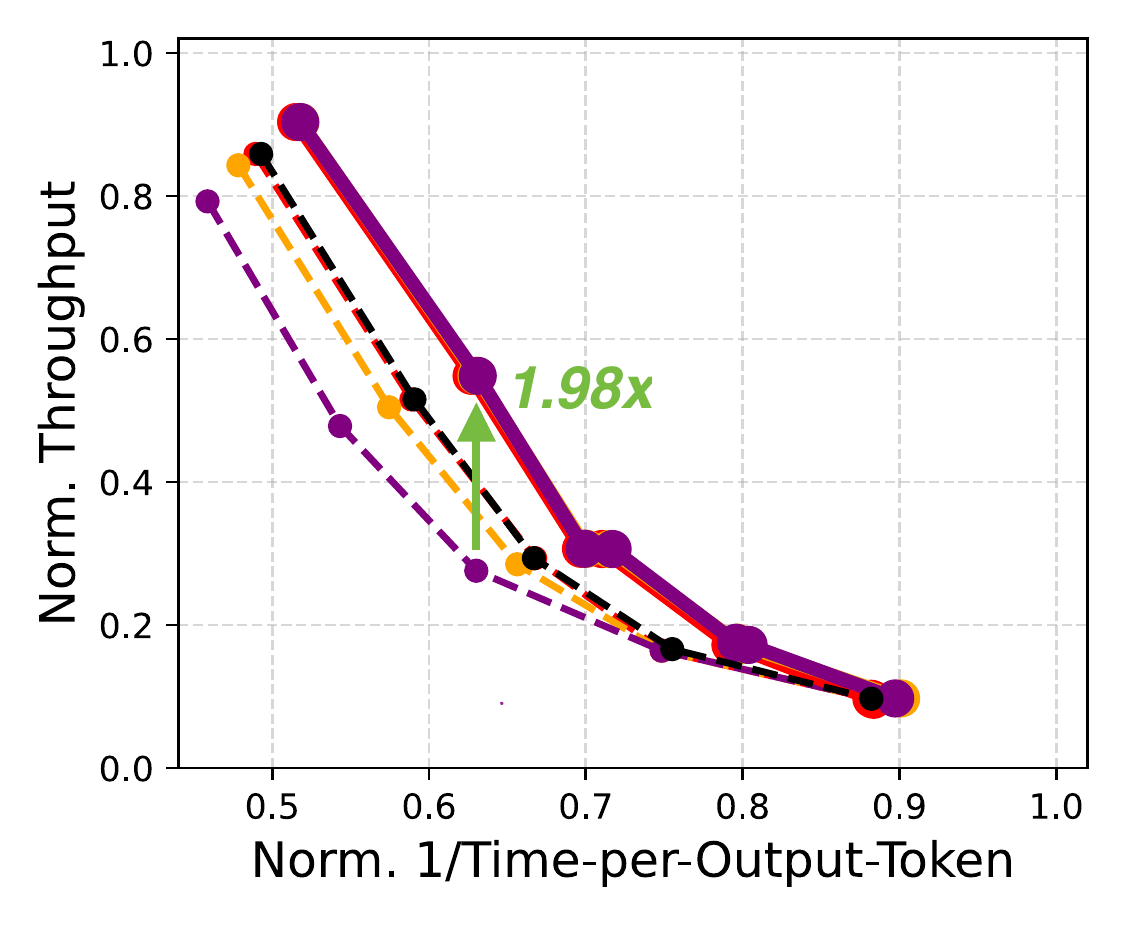}\label{fig:decode_pareto_curve_qwen_gsm8k}}
    \hfill
  \subfloat[DeepSeekV3, Humaneval, 16*B200]{\includegraphics[width=0.248\textwidth]{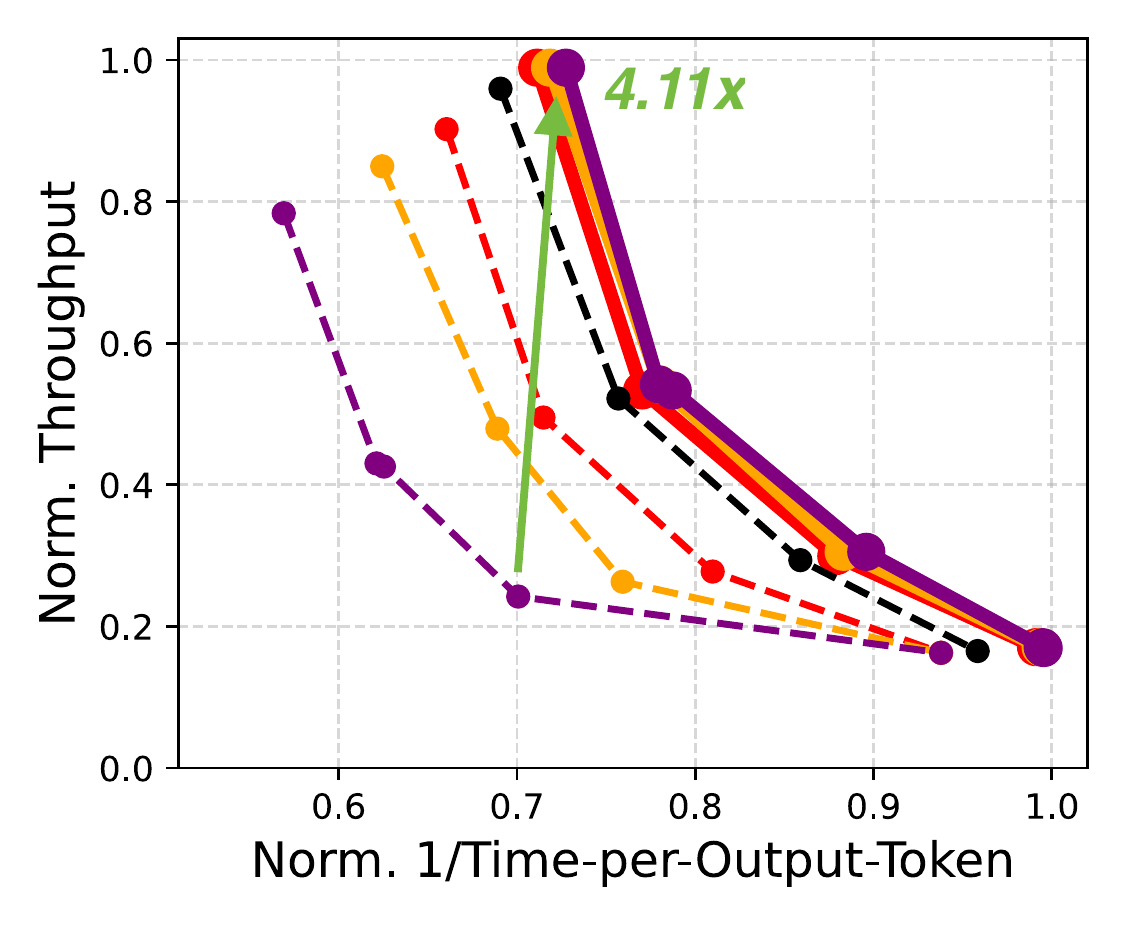}\label{fig:decode_pareto_curve_deepseek_humaneval}}
  \hfill
    \subfloat[DeepSeekV3, GSM8K, 16*B200]{\includegraphics[width=0.248\textwidth]{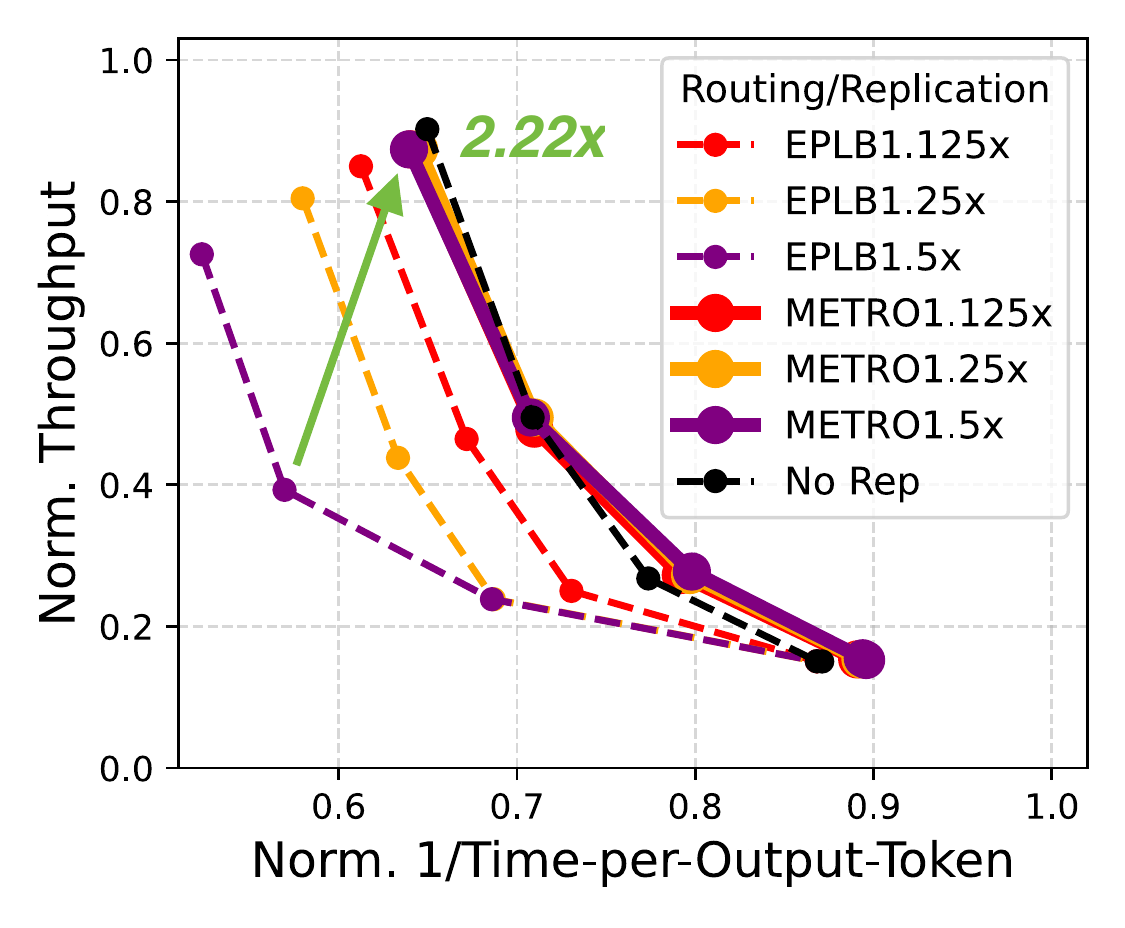}\label{fig:decode_pareto_curve_deepseek_gsm8k}}
    \hfill
  \caption{\textbf{Pareto curves of the decode phase with simulation setup in \S\ref{ssec:system_setup} and Table~\ref{tab:decode_pareto_dlsim_config}.} Both the x-axis and the y-axis are normalized to the maximum attainable values due to confidentiality. For a fixed TPOT (representing a specific SLO), \name delivers remarkably higher decode throughput of $1.98\times$ -- $4.11\times$ across models and datasets. As the throughput and TPOT objectives change, \name consistently outperforms EPLB routing across different replication ratios, batch sizes, and TP/EP mappings for both models and datasets. Higher replication ratio improves \name's performance gain because it exacerbates EPLB's inflation on activated experts.
  }
  \label{fig:decode_pareto_curve}
\end{figure*}

\begin{tcolorbox}
\emph{Takeaway}: with EPLB routing, while prefill-heavy workloads benefit from higher replication, decode-heavy workloads fare poorly under the same high replication factors. \name eliminates the need for workload-specific calibration, delivering consistently higher performance than the EPLB routing (up to $21.0\%$ for decode-heavy and up to $4.2\%$ for prefill-heavy workloads). 
\end{tcolorbox}

\paragraphb{Latency Breakdown} 
Fig.~\ref{fig:design_latency_breakdown} shows the breakdown for one layer of Qwen3-30B with various replication ratios on our real system implementation. All three sources of \name's performance overheads (detailed in \S\ref{ssec:metro_algorithm}) are insignificant compared to its reduction on FFN latency. First, executing Algorithm~\ref{ssec:exact_algorithm} adds up to $26\mu s$ of latency for routing at a replication ratio of $1.5\times$, but this is more than offset by the $81 \mu s$ reduction in FFN time. Second, top-$k$ originally accounts for only $17 \mu s$--$19 \mu s$ ($<5\%$ of the layer time), and extending it to all tokens adds at most $3 \mu s$ ($<1\%$). Lastly, we observe no statistically significant increase in \allgather's communication time compared to \alltoall. As explained in \S\ref{ssec:exact_algorithm}, this is because in the memory-bound regime (\ie, small batches), NVLink latency dominates and bandwidth cost is far below the fixed cost of launching NCCL collectives.

\subsection{Decode Throughput-Latency Pareto-optimality Analysis}
\label{ssec:decode_only_analysis}

While our earlier evaluations focus on a single, representative decode configuration (\ie, full EP, global batch size = $1K$), real-world deployments may require varying service-level objectives (SLOs) which demand a broader set of deployment configurations.
 Specifically, there exists a fundamental tradeoff between decode throughput and latency (Time-per-Output-Token, TPOT). Different points along the tradeoff space may prefer different batch sizes or different model parallelisms. To efficiently navigate through these configurations, we conduct a decode-only analysis (since \name does not affect prefill performance). We simulated all batch sizes and parallelism configurations listed in Table~\ref{tab:decode_pareto_dlsim_config} across various replication ratios, and visualized the Pareto-optimal configuration curve for decode throughput and latency (\ie, TPOT) in Fig.~\ref{fig:decode_pareto_curve}. All simulations use the same hardware and dataset configurations described in \S\ref{ssec:system_setup}. We focus on understanding two research questions:
 \begin{itemize}
    \item \textbf{Q1}: How does \name improve the decode throughput-latency Pareto frontier compared to EPLB?
    \item \textbf{Q2}: How does prioritizing higher decode throughput vs. lower decode latency impact the choice of deployment configuration and, in turn, the benefits of \name?
  \end{itemize}



\paragraphb{Q1: \name's impact on the Pareto frontier} Fig.~\ref{fig:decode_pareto_curve} shows that for a fixed TPOT (representing a specific SLO), \name delivers remarkably higher decode throughput of $1.98\times$ -- $4.11\times$ compared to EPLB across models and datasets. This throughput improvement exceeds that in our earlier evaluation (\S\ref{ssec:endtoend_perf}) since more flexible batch-size choices coupled with \name’s latency reduction enable the use of larger batches to increase throughput while still meeting request SLOs. Notably, the highest throughput improvement is achieved with DeepSeek-V3 and Humaneval~(\ie, Fig.~\ref{fig:decode_pareto_curve_deepseek_humaneval}) at $1.5\times$ replication and normalized 1/TPOT = $0.7$, because \name's reduced TPOT allows it to use $4\times$ larger batch size compared to EPLB with the same request SLOs.

Moreover, we observe that higher expert replication ratios amplify \name's performance gain over EPLB (a similar observation was discussed in \S\ref{ssec:endtoend_perf}). This is because with more replication, EPLB increases the number of activated experts across all configurations. In contrast, the number of activated experts under \name remains the same or even reduces (\ie, Humaneval, Fig.~\ref{fig:decode_pareto_curve_qwen_humaneval} and Fig.~\ref{fig:decode_pareto_curve_deepseek_humaneval}) with more replication.

Note that while no-replication yields comparable performance to \name in the decode phase, it leads to significantly higher latency for the prefill phase (\ie, by up to $38\%$ compared to $1.5x$ replication), potentially violating SLOs and degrading overall token throughput.

\begin{tcolorbox}
\emph{Takeaway}: \name consistently advances the decode throughput-latency Pareto frontier compared to EPLB across datasets, models, and expert replication ratios, achieving $1.98\times$ -- $4.11\times$ higher decode throughput under the same SLO with EPLB.
\end{tcolorbox}



\begin{table}[t]
\centering
\caption{Decode Pareto Curve Simulation Setup}
\label{tab:decode_pareto_dlsim_config}
\begin{tabular}{lcc}
\hline
Models & Qwen3-235B-A22B & DeepSeek-V3-671B \\
\hline
Hardware & 8*B200 & 16*B200 \\
\hline
Batch Sizes & 1024/512/256/128/64& 1024/512/256/128 \\
\hline
Parallelism & TP{1,2,4,8} * EP{1,2,4,8} & TP{1,2,4,8,16} * EP{1,2,4,8,16} \\
\hline
\end{tabular}
\end{table}

\begin{figure}[t!]
  \centering
 \includegraphics[width=0.32\textwidth]{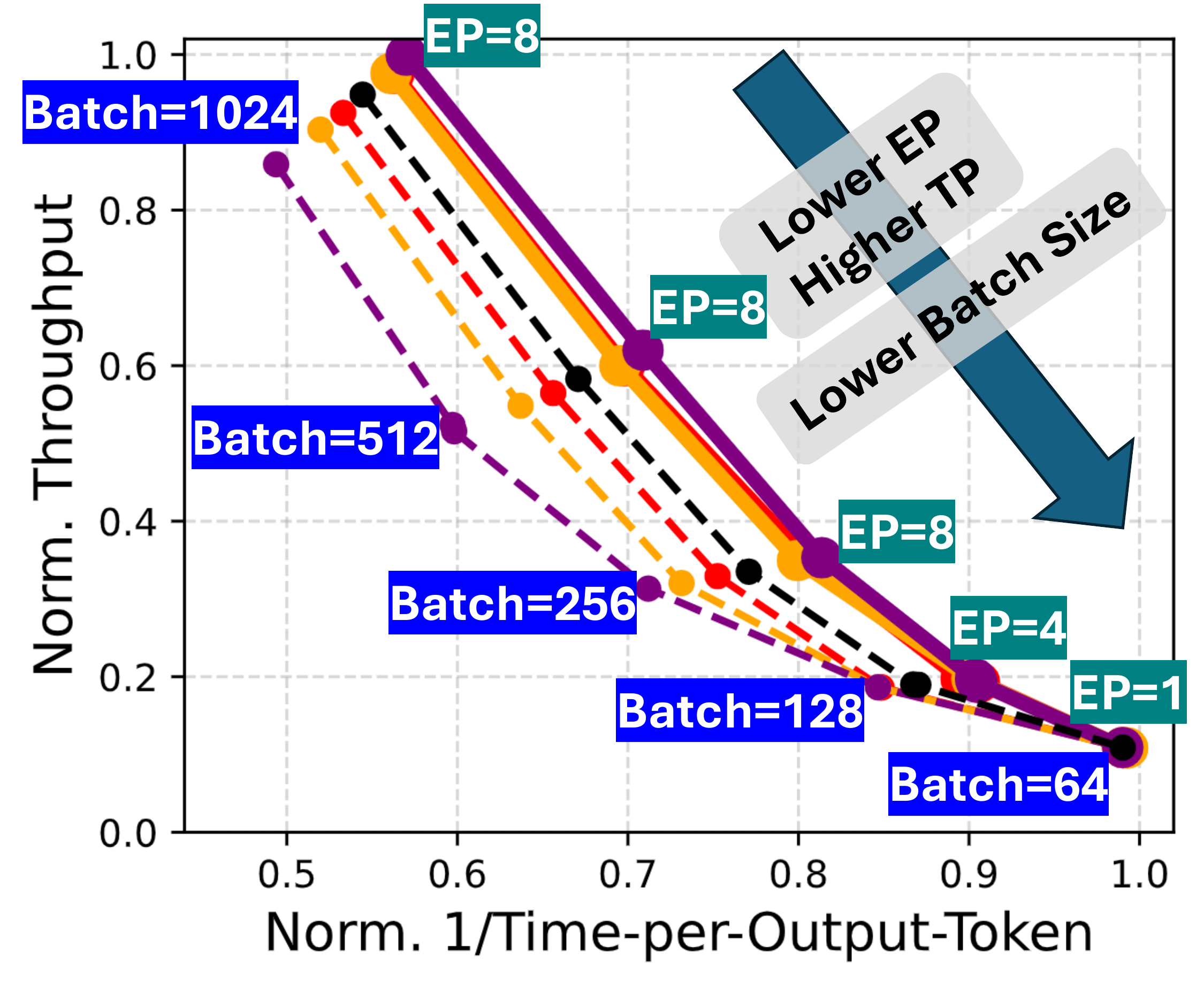}
  \caption{\textbf{
  Annotated Pareto curve analysis using Fig.~\ref{fig:decode_pareto_curve_qwen_humaneval} as an example (Q2).} Other curves shown in Fig.~\ref{fig:decode_pareto_curve} follow similar trends. Batch size decreases when moving from high-throughput to low latency. With extremely low latency requirements, the batch size is small enough that network latency becomes the bottleneck, where TP is preferred because its communication overhead is minimal, making EP load balancing unnecessary.}\label{fig:ep_tp_mapping_network_latency1}
\end{figure}


\paragraphb{Q2: SLO's impact on \name's performance gain}

We observe that when an extremely low TPOT (\ie, strict SLO) is required, \name's performance gains can be diminished. Specifically, as shown in Fig.~\ref{fig:ep_tp_mapping_network_latency1}, achieving lower TPOT requires smaller batch sizes to ensure that tokens are processed sooner. When the required TPOT is extremely low~(\ie, 1/TPOT $>0.9$), the batch size is small enough (\ie, $\leq 64$) to cause the workload to \textit{no longer be memory-bound}, making the inter-GPU network latency the dominant bottleneck. In such scenarios, maximizing TP (rather than EP) is preferable, as smaller batch sizes require much lower communication bandwidth under TP --- which is typically constrained by network bandwidth bottlenecks. Since a full TP solution eliminates any load imbalance across GPUs that may degrade performance, all curves converge to full TP at the bottom-right of Fig.~\ref{fig:ep_tp_mapping_network_latency1}, which makes any EP load balancing unnecessary. In addition, we do not observe any Pareto-optimal configurations using pipeline-parallelism. This is because of its degradation on decode latency (\ie, by a factor of pipeline stages) outweighs its potential improvement on throughput.


\begin{tcolorbox}
\emph{Takeaway}: {\name can benefit major SLO regimes of the decode phase; in extreme cases, low TPOT (strict SLO) requires a smaller batch size, making network latency the bottleneck instead of memory bandwidth, where full TP delivers the best performance, eliminating the need for any EP load balancing schemes.}
\end{tcolorbox} 

%% file: tex/discussion.tex
\section{Discussion and Future Work}\label{sec:discussion}

We now discuss \name's applicability to emerging disaggregated prefill-decode deployments and to future models, workloads, and accelerator hardware.

\subsection{Prefill-decode disaggregation}
Recent works have argued for disaggregating prefill and decode phases~\cite{agrawal2024sarathi_serve, fu2024serverlessllm, gao2024cachedattention, xupeng2024specinfer, pratyush2024splitwise, zhang2025spad}, a deployment strategy that executes prefill and decode on separate GPUs in a multi-GPU system instead of co-locating them on all GPUs, where EP load balancing for the compute-bound phase (\ie, prefill) and the memory-bound phase (\ie, decode) can be decoupled. Under disaggregation, while the decode phase would favor no-replication than EPLB with higher replication ratio (Fig.~\ref{fig:decode_pareto_curve}), \name can still improve on no-replication in most setups --- by up to 4.3\% with $1.125\times$ replication and up to 5.0\% with $1.5\times$ replication as shown in Fig.~\ref{fig:decode_pareto_curve}.

On the other hand, we note that \name's major benefits still lies in prefill-decode co-deployed settings as we evaluated in \S\ref{sec:eval}, a deployment strategy more commonly seen in smaller systems~\cite{agrawal2024sarathi_serve, fu2024serverlessllm, xupeng2024specinfer, gao2024cachedattention, aditya2025podattention, zhu2025nanoflow} --- e.g., a single server with a few GPUs.
We anticipate that such prefill-decode co-deployment will remain popular for small systems in the future. This is due to disaggregation's various inefficiencies for small systems that may outweigh its benefit, including the communication overheads to transfer KV cache between the prefill and decode instances~\cite{zhang2025spad, gao2024cachedattention}, the memory capacity pressure stemming from duplicating model weights across the two instances~\cite{zhang2025spad}, and the need to dynamically re-provision resources across the two phases under workloads with unpredictable prefill-decode token ratios~\cite{agrawal2024sarathi_serve}. As such, we anticipate that \name will remain beneficial to small multi-GPU systems' MoE deployment in the future.


\subsection{Future Models, Workloads and Hardware}

Looking ahead, we anticipate a similar, if not greater, need for memory-efficient expert routing schemes. With increasingly larger MoE models, future workloads are likely to be more memory bandwidth limited due to ballooning model weights and sparser activation patterns~\cite{moe_survey1, deepseekai2025deepseekv3technicalreport, qwen3technicalreport}. From the hardware perspective, although new generations of GPUs offer higher compute capability and memory bandwidth, the increase in bandwidth might not keep pace with compute; memory remains expensive, and its linear growth seems to be lagging behind compute improvements~\cite{digitalocean-gpu-memory-bandwidth, davies2025efficientllminferencebandwidth, nvidia_a100_whitepaper_2020, nvidia_h100_whitepaper_2022, nvidia_blackwell_tech_overview_2025}. As such, we believe that optimizing for memory bandwidth will remain a critical challenge. Since \name is designed to minimize memory traffic during expert routing, it will retain its benefits in terms of lower decode latency and higher overall throughput for future workloads.

%% file: tex/conclusion.tex
\section{Conclusion}
\label{sec:conclusion}

Existing load-balancing approaches for expert-parallel MoE serving aim to balance the number of tokens each GPU processes. We show that this approach \textit{degrades} performance rather than improving it when inference is memory-bound due to the inflated number of activated experts that exacerbates memory pressure.
We propose \name, a novel token-routing algorithm for high-performance expert-parallel MoE serving in the memory-bound regime that minimizes the number of activated experts per GPU rather than balancing tokens. Our evaluation of \name against EPLB on both real systems (vLLM over 8 A100 GPUs) and a proprietary simulator (8-16 B200 GPUs) shows that \name reduces decode latency by $11$ - $22\%$, and total token throughput by $3$ - $21\%$ for Qwen3 and DeepSeek-V3 serving, where prefill and decode phases are co-deployed. In addition, by trading latency headroom for throughput, \name improves decode throughput by up to $4.11\times$ over EPLB at a fixed decode SLO.

%% file: bib/abr-short.bib
@STRING{ASPLOS = "Proc. ACM ASPLOS"}

@STRING{ISCA = "Proc. IEEE ISCA"}

@STRING{OSDI = "Proc. USENIX OSDI"}

@STRING{PPOPP = "Proc. ACM PPoPP"}

@string{ATC = "Proc. USENIX ATC"}

@STRING{MLSYS = "Proc. MLSys"}

@STRING{NeurIPS      = "Proc. NeurIPS"}

@STRING{ISCA="Proc. ACM/IEEE ISCA"}

@STRING{IPDPS="Proc. IEEE IPDPS"}

@STRING{FAST = "Proc. USENIX FAST"}


%% file: bib/paper.bib
@misc{eplb-repo,
  author       = {Shaoyuan Chen and A-transformer and haosdent and Chun Chen},
  title        = {{EPLB}: Expert Parallelism Load Balancer},
  howpublished = {\url{https://github.com/deepseek-ai/EPLB}},
  year         = {2025},
}

@article{flexmoe,
    author = {Nie, Xiaonan and Miao, Xupeng and Wang, Zilong and Yang, Zichao and Xue, Jilong and Ma, Lingxiao and Cao, Gang and Cui, Bin},
    title = {FlexMoE: Scaling Large-scale Sparse Pre-trained Model Training via Dynamic Device Placement},
    year = {2023},
    journal = {Proc. ACM Manag. Data},
}

@ARTICLE{moesys,
  author={Yu, Dianhai and Shen, Liang and Hao, Hongxiang and Gong, Weibao and Wu, Huachao and Bian, Jiang and Dai, Lirong and Xiong, Haoyi},
  journal={IEEE Transactions on Services Computing}, 
  title={MoESys: A Distributed and Efficient Mixture-of-Experts Training and Inference System for Internet Services}, 
  year={2024},
}

@inproceedings{fastermoe,
    author = {He, Jiaao and Zhai, Jidong and Antunes, Tiago and Wang, Haojie and Luo, Fuwen and Shi, Shangfeng and Li, Qin},
    title = {FasterMoE: modeling and optimizing training of large-scale dynamic pre-trained models},
    year = {2022},
    booktitle = PPOPP
}

@inproceedings {smartmoe,
    author = {Mingshu Zhai and Jiaao He and Zixuan Ma and Zan Zong and Runqing Zhang and Jidong Zhai},
    title = {{SmartMoE}: Efficiently Training {Sparsely-Activated} Models through Combining Offline and Online Parallelization},
    booktitle = ATC,
    year = {2023},
}

@article{vllm,
  title={Efficient Memory Management for Large Language Model Inference},
  author={Kwon, Woosuk and Li, Zhuohan and Zhuang, Siyuan and Sheng, Ying and Zheng, Lianmin and Yu, Cody and Gonzalez, Joseph E and Zhang, Hao and Stoica, Ion},
  journal={arXiv preprint arXiv:2309.06180},
  year={2023}
}

@inproceedings{sglang,
    author = {Zheng, Lianmin and Yin, Liangsheng and Xie, Zhiqiang and Sun, Chuyue and Huang, Jeff and Yu, Cody Hao and Cao, Shiyi and Kozyrakis, Christos and Stoica, Ion and Gonzalez, Joseph E. and Barrett, Clark and Sheng, Ying},
    title = {SGLang: efficient execution of structured language model programs},
    year = {2025},
    booktitle = NeurIPS
}

@ARTICLE{moe_survey1,
    author={Cai, Weilin and Jiang, Juyong and Wang, Fan and Tang, Jing and Kim, Sunghun and Huang, Jiayi},
    journal={ IEEE Transactions on Knowledge \& Data Engineering },
    title={{ A Survey on Mixture of Experts in Large Language Models }},
    year={2025}
}

@misc{davies2025efficientllminferencebandwidth,
      title={Efficient LLM Inference: Bandwidth, Compute, Synchronization, and Capacity are all you need}, 
      author={Michael Davies and Neal Crago and Karthikeyan Sankaralingam and Christos Kozyrakis},
      year={2025},
      eprint={2507.14397},
      archivePrefix={arXiv},
      primaryClass={cs.AR},
      url={https://arxiv.org/abs/2507.14397}, 
}

@article{gemini15report,
  title   = {Gemini 1.5: Unlocking Multimodal Understanding Across Millions of Tokens of Context},
  author  = {{Gemini Team, Google}},
  journal = {arXiv preprint arXiv:2403.05530},
  year    = {2024},
  url     = {https://arxiv.org/abs/2403.05530}
}

@misc{anthropic_sonnet4_1m_2025,
  author       = {{Anthropic}},
  title        = {Claude Sonnet 4 now supports 1M tokens of context},
  howpublished = {\url{https://www.anthropic.com/news/1m-context}},
  year         = {2025}
}

@misc{zhang2025tensorcoresbenefitmemorybound,
      title={Can Tensor Cores Benefit Memory-Bound Kernels? (No!)}, 
      author={Lingqi Zhang and Jiajun Huang and Sheng Di and Satoshi Matsuoka and Mohamed Wahib},
      year={2025},
      eprint={2502.16851},
      archivePrefix={arXiv},
      primaryClass={cs.DC},
      url={https://arxiv.org/abs/2502.16851}, 
}

@misc{yuan2024llminferenceunveiledsurvey,
      title={LLM Inference Unveiled: Survey and Roofline Model Insights}, 
      author={Zhihang Yuan and Yuzhang Shang and Yang Zhou and Zhen Dong and Zhe Zhou and Chenhao Xue and Bingzhe Wu and Zhikai Li and Qingyi Gu and Yong Jae Lee and Yan Yan and Beidi Chen and Guangyu Sun and Kurt Keutzer},
      year={2024},
      eprint={2402.16363},
      archivePrefix={arXiv},
      primaryClass={cs.CL},
      url={https://arxiv.org/abs/2402.16363}, 
}

@inproceedings{agrawal2024sarathi_serve,
    author = {Agrawal, Amey and Kedia, Nitin and Panwar, Ashish and Mohan, Jayashree and Kwatra, Nipun and Gulavani, Bhargav S. and Tumanov, Alexey and Ramjee, Ramachandran},
    title = {Taming throughput-latency tradeoff in LLM inference with sarathi-serve},
    year = {2024},
    booktitle = OSDI
}

@misc{deepseekai2025deepseekv3technicalreport,
      title={DeepSeek-V3 Technical Report}, 
      author={DeepSeek-AI and Aixin Liu and Bei Feng and Bing Xue and Bingxuan Wang and Bochao Wu and Chengda Lu and Chenggang Zhao and Chengqi Deng and Chenyu Zhang and Chong Ruan and Damai Dai and Daya Guo and Dejian Yang and Deli Chen and Dongjie Ji and Erhang Li and Fangyun Lin and Fucong Dai and Fuli Luo and Guangbo Hao and Guanting Chen and Guowei Li and H. Zhang and Han Bao and Hanwei Xu and Haocheng Wang and Haowei Zhang and Honghui Ding and Huajian Xin and Huazuo Gao and Hui Li and Hui Qu and J. L. Cai and Jian Liang and Jianzhong Guo and Jiaqi Ni and Jiashi Li and Jiawei Wang and Jin Chen and Jingchang Chen and Jingyang Yuan and Junjie Qiu and Junlong Li and Junxiao Song and Kai Dong and Kai Hu and Kaige Gao and Kang Guan and Kexin Huang and Kuai Yu and Lean Wang and Lecong Zhang and Lei Xu and Leyi Xia and Liang Zhao and Litong Wang and Liyue Zhang and Meng Li and Miaojun Wang and Mingchuan Zhang and Minghua Zhang and Minghui Tang and Mingming Li and Ning Tian and Panpan Huang and Peiyi Wang and Peng Zhang and Qiancheng Wang and Qihao Zhu and Qinyu Chen and Qiushi Du and R. J. Chen and R. L. Jin and Ruiqi Ge and Ruisong Zhang and Ruizhe Pan and Runji Wang and Runxin Xu and Ruoyu Zhang and Ruyi Chen and S. S. Li and Shanghao Lu and Shangyan Zhou and Shanhuang Chen and Shaoqing Wu and Shengfeng Ye and Shengfeng Ye and Shirong Ma and Shiyu Wang and Shuang Zhou and Shuiping Yu and Shunfeng Zhou and Shuting Pan and T. Wang and Tao Yun and Tian Pei and Tianyu Sun and W. L. Xiao and Wangding Zeng and Wanjia Zhao and Wei An and Wen Liu and Wenfeng Liang and Wenjun Gao and Wenqin Yu and Wentao Zhang and X. Q. Li and Xiangyue Jin and Xianzu Wang and Xiao Bi and Xiaodong Liu and Xiaohan Wang and Xiaojin Shen and Xiaokang Chen and Xiaokang Zhang and Xiaosha Chen and Xiaotao Nie and Xiaowen Sun and Xiaoxiang Wang and Xin Cheng and Xin Liu and Xin Xie and Xingchao Liu and Xingkai Yu and Xinnan Song and Xinxia Shan and Xinyi Zhou and Xinyu Yang and Xinyuan Li and Xuecheng Su and Xuheng Lin and Y. K. Li and Y. Q. Wang and Y. X. Wei and Y. X. Zhu and Yang Zhang and Yanhong Xu and Yanhong Xu and Yanping Huang and Yao Li and Yao Zhao and Yaofeng Sun and Yaohui Li and Yaohui Wang and Yi Yu and Yi Zheng and Yichao Zhang and Yifan Shi and Yiliang Xiong and Ying He and Ying Tang and Yishi Piao and Yisong Wang and Yixuan Tan and Yiyang Ma and Yiyuan Liu and Yongqiang Guo and Yu Wu and Yuan Ou and Yuchen Zhu and Yuduan Wang and Yue Gong and Yuheng Zou and Yujia He and Yukun Zha and Yunfan Xiong and Yunxian Ma and Yuting Yan and Yuxiang Luo and Yuxiang You and Yuxuan Liu and Yuyang Zhou and Z. F. Wu and Z. Z. Ren and Zehui Ren and Zhangli Sha and Zhe Fu and Zhean Xu and Zhen Huang and Zhen Zhang and Zhenda Xie and Zhengyan Zhang and Zhewen Hao and Zhibin Gou and Zhicheng Ma and Zhigang Yan and Zhihong Shao and Zhipeng Xu and Zhiyu Wu and Zhongyu Zhang and Zhuoshu Li and Zihui Gu and Zijia Zhu and Zijun Liu and Zilin Li and Ziwei Xie and Ziyang Song and Ziyi Gao and Zizheng Pan},
      year={2025},
      eprint={2412.19437},
      archivePrefix={arXiv},
      primaryClass={cs.CL},
      url={https://arxiv.org/abs/2412.19437}, 
}

@misc{qwen3technicalreport,
      title={Qwen3 Technical Report}, 
      author={Qwen Team},
      year={2025},
      eprint={2505.09388},
      archivePrefix={arXiv},
      primaryClass={cs.CL},
      url={https://arxiv.org/abs/2505.09388}, 
}

@inproceedings{huang2024mindthegap,
    author = {Huang, Qijing and Tsai, Po-An and Emer, Joel S. and Parashar, Angshuman},
    title = {Mind the Gap: Attainable Data Movement and Operational Intensity Bounds for Tensor Algorithms},
    year = {2025},
    booktitle = ISCA
}

@misc{nvidia_h100_whitepaper_2022,
  author       = {{NVIDIA}},
  title        = {{NVIDIA H100 Tensor Core GPU Architecture}},
  howpublished = {\url{https://www.advancedclustering.com/wp-content/uploads/2022/03/gtc22-whitepaper-hopper.pdf}},
  year         = {2022},
  note         = {H100 (Hopper) architecture whitepaper; Accessed: 2025-11-02}
}

@misc{nvidia_a100_whitepaper_2020,
  author       = {{NVIDIA}},
  title        = {{NVIDIA A100 Tensor Core GPU Architecture}},
  howpublished = {\url{http://images.nvidia.com/aem-dam/en-zz/Solutions/data-center/nvidia-ampere-architecture-whitepaper.pdf}},
  year         = {2020},
}

@misc{nvidia_blackwell_tech_overview_2025,
  author       = {{NVIDIA}},
  title        = {{NVIDIA Blackwell Architecture Technical Overview}},
  howpublished = {\url{https://resources.nvidia.com/en-us-blackwell-architecture}},
  year         = {2025},
  note         = {Blackwell generation overview including B200; Accessed: 2025-11-02}
}

@misc{li2024instructcoderinstructiontuninglarge,
      title={InstructCoder: Instruction Tuning Large Language Models for Code Editing}, 
      author={Kaixin Li and Qisheng Hu and Xu Zhao and Hui Chen and Yuxi Xie and Tiedong Liu and Qizhe Xie and Junxian He},
      year={2024},
      eprint={2310.20329},
      archivePrefix={arXiv},
      primaryClass={cs.CL},
      url={https://arxiv.org/abs/2310.20329}, 
}

@misc{cheng2025barbariansgateaiupending,
      title={Barbarians at the Gate: How AI is Upending Systems Research}, 
      author={Audrey Cheng and Shu Liu and Melissa Pan and Zhifei Li and Bowen Wang and Alex Krentsel and Tian Xia and Mert Cemri and Jongseok Park and Shuo Yang and Jeff Chen and Lakshya Agrawal and Aditya Desai and Jiarong Xing and Koushik Sen and Matei Zaharia and Ion Stoica},
      year={2025},
      eprint={2510.06189},
      archivePrefix={arXiv},
      primaryClass={cs.AI},
      url={https://arxiv.org/abs/2510.06189}, 
}

@inproceedings{dinic1970algorithm,
  title={Algorithm for solution of a problem of maximum flow in networks with power estimation},
  author={Dinic, Efim A},
  booktitle={Soviet Math. Doklady},
  year={1970}
}

@misc{exact_algo_1,
      title={The 2-valued case of makespan minimization with assignment constraints}, 
      author={Stavros G. Kolliopoulos and Yannis Moysoglou},
      year={2012},
      eprint={1212.1609},
      archivePrefix={arXiv},
      primaryClass={cs.DS},
      url={https://arxiv.org/abs/1212.1609}, 
}

@article{exact_algo_2,
	author = {Chung-Lun Li},
	journal = {European Journal of Operational Research},
	title = {Scheduling unit-length jobs with machine eligibility restrictions},
	year = {2006}
}

@article{exact_algo_3,
  title={Scheduling identical jobs on uniform parallel machines},
  author={Dessouky, Mohamed I and Lageweg, Ben J and Lenstra, Jan Karel and van de Velde, Steef L},
  journal={Statistica Neerlandica},
  year={1990},
}

@inproceedings{vanausdalefficient,
  title={An Efficient Push-Relabel Implementation for Max-Flow Computations on GPUs},
  author={VanAusdal, Avery and Burtscher, Martin},
    booktitle = {Proc. IEEE International Performance Computing and Communications Conference},
    year = 2025
}

@misc{vllm-cuda-graphs,
  title        = {vLLM Documentation: CUDA Graphs},
  howpublished = {\url{https://docs.vllm.ai/en/latest/design/cuda_graphs/}},
}

@misc{taslock,
  title = {Test-And-Set lock},
  howpublished = "\url{https://www.cs.rochester.edu/research/synchronization/pseudocode/ss.html#tas}"
}

@misc{fedus2022switchtransformersscalingtrillion,
      title={Switch Transformers: Scaling to Trillion Parameter Models with Simple and Efficient Sparsity}, 
      author={William Fedus and Barret Zoph and Noam Shazeer},
      year={2022},
      eprint={2101.03961},
      archivePrefix={arXiv},
      primaryClass={cs.LG},
      url={https://arxiv.org/abs/2101.03961}, 
}

@misc{lepikhin2020gshardscalinggiantmodels,
      title={GShard: Scaling Giant Models with Conditional Computation and Automatic Sharding}, 
      author={Dmitry Lepikhin and HyoukJoong Lee and Yuanzhong Xu and Dehao Chen and Orhan Firat and Yanping Huang and Maxim Krikun and Noam Shazeer and Zhifeng Chen},
      year={2020},
      eprint={2006.16668},
      archivePrefix={arXiv},
      primaryClass={cs.CL},
      url={https://arxiv.org/abs/2006.16668}, 
}

@misc{shazeer2017outrageouslylargeneuralnetworks,
      title={Outrageously Large Neural Networks: The Sparsely-Gated Mixture-of-Experts Layer}, 
      author={Noam Shazeer and Azalia Mirhoseini and Krzysztof Maziarz and Andy Davis and Quoc Le and Geoffrey Hinton and Jeff Dean},
      year={2017},
      eprint={1701.06538},
      archivePrefix={arXiv},
      primaryClass={cs.LG},
      url={https://arxiv.org/abs/1701.06538}, 
}

@article{Williams2009Roofline,
    author = {Williams, Samuel and Waterman, Andrew and Patterson, David},
    title = {Roofline: an insightful visual performance model for multicore architectures},
    year = {2009},
    journal = {Commun. ACM}
}

@inproceedings{zhong2024distserve,
    author = {Zhong, Yinmin and Liu, Shengyu and Chen, Junda and Hu, Jianbo and Zhu, Yibo and Liu, Xuanzhe and Jin, Xin and Zhang, Hao},
    title = {DistServe: disaggregating prefill and decoding for goodput-optimized large language model serving},
    year = {2024},
    booktitle = OSDI
}

@inproceedings{pratyush2024splitwise,
    author = {Patel, Pratyush and Choukse, Esha and Zhang, Chaojie and Shah, Aashaka and Goiri, \'{I}\~{n}igo and Maleki, Saeed and Bianchini, Ricardo},
    title = {Splitwise: Efficient Generative LLM Inference Using Phase Splitting},
    year = {2025},
    booktitle = ISCA,
}

@misc{nvidia2025dynamo,
  title        = {NVIDIA Dynamo: A Low-Latency Distributed Inference Framework for Scaling Reasoning {AI} Models},
  howpublished = {NVIDIA Developer Blog},
  year         = {2025},
  url          = {https://developer.nvidia.com/dynamo}
}

@inproceedings {qin2025mooncake,
    author = {Ruoyu Qin and Zheming Li and Weiran He and Jialei Cui and Feng Ren and Mingxing Zhang and Yongwei Wu and Weimin Zheng and Xinran Xu},
    title = {Mooncake: Trading More Storage for Less Computation {\textemdash} A {KVCache-centric} Architecture for Serving {LLM} Chatbot},
    booktitle = FAST,
    year = {2025},
}

@inproceedings {fu2024serverlessllm,
    author = {Yao Fu and Leyang Xue and Yeqi Huang and Andrei-Octavian Brabete and Dmitrii Ustiugov and Yuvraj Patel and Luo Mai},
    title = {{ServerlessLLM}: {Low-Latency} Serverless Inference for Large Language Models},
    booktitle = OSDI,
    year = {2024}
}

@inproceedings{xupeng2024specinfer,
    author = {Miao, Xupeng and Oliaro, Gabriele and Zhang, Zhihao and Cheng, Xinhao and Wang, Zeyu and Zhang, Zhengxin and Wong, Rae Ying Yee and Zhu, Alan and Yang, Lijie and Shi, Xiaoxiang and Shi, Chunan and Chen, Zhuoming and Arfeen, Daiyaan and Abhyankar, Reyna and Jia, Zhihao},
    title = {SpecInfer: Accelerating Large Language Model Serving with Tree-based Speculative Inference and Verification},
    year = {2024},
    booktitle = ASPLOS
}

@inproceedings {gao2024cachedattention,
    author = {Bin Gao and Zhuomin He and Puru Sharma and Qingxuan Kang and Djordje Jevdjic and Junbo Deng and Xingkun Yang and Zhou Yu and Pengfei Zuo},
    title = {{Cost-Efficient} Large Language Model Serving for Multi-turn Conversations with {CachedAttention}},
    booktitle = ATC,
    year = {2024}
}

@article{cobbe2021gsm8k,
  title={Training Verifiers to Solve Math Word Problems},
  author={Cobbe, Karl and Kosaraju, Vineet and Bavarian, Mohammad and Chen, Mark and Jun, Heewoo and Kaiser, Lukasz and Plappert, Matthias and Tworek, Jerry and Hilton, Jacob and Nakano, Reiichiro and Hesse, Christopher and Schulman, John},
  journal={arXiv preprint arXiv:2110.14168},
  year={2021}
}

@misc{chen2021evaluating,
      title={Evaluating Large Language Models Trained on Code},
      author={Mark Chen and Jerry Tworek and Heewoo Jun and Qiming Yuan and Henrique Ponde de Oliveira Pinto and Jared Kaplan and Harri Edwards and Yuri Burda and Nicholas Joseph and Greg Brockman and Alex Ray and Raul Puri and Gretchen Krueger and Michael Petrov and Heidy Khlaaf and Girish Sastry and Pamela Mishkin and Brooke Chan and Scott Gray and Nick Ryder and Mikhail Pavlov and Alethea Power and Lukasz Kaiser and Mohammad Bavarian and Clemens Winter and Philippe Tillet and Felipe Petroski Such and Dave Cummings and Matthias Plappert and Fotios Chantzis and Elizabeth Barnes and Ariel Herbert-Voss and William Hebgen Guss and Alex Nichol and Alex Paino and Nikolas Tezak and Jie Tang and Igor Babuschkin and Suchir Balaji and Shantanu Jain and William Saunders and Christopher Hesse and Andrew N. Carr and Jan Leike and Josh Achiam and Vedant Misra and Evan Morikawa and Alec Radford and Matthew Knight and Miles Brundage and Mira Murati and Katie Mayer and Peter Welinder and Bob McGrew and Dario Amodei and Sam McCandlish and Ilya Sutskever and Wojciech Zaremba},
      year={2021},
      eprint={2107.03374},
      archivePrefix={arXiv},
      primaryClass={cs.LG}
}

@online{nvidia_dgx_b200_web,
  title = {NVIDIA DGX B200 - The foundation for your AI factory},
  author = {NVIDIA Corporation},
  year = {2025},
  url = {https://www.nvidia.com/en-us/data-center/dgx-b200/},
  note = {Product specifications and overview for the NVIDIA DGX B200 system.}
}

@misc{numina_math_datasets,
  author = {Jia LI and Edward Beeching and Lewis Tunstall and Ben Lipkin and Roman Soletskyi and Shengyi Costa Huang and Kashif Rasul and Longhui Yu and Albert Jiang and Ziju Shen and Zihan Qin and Bin Dong and Li Zhou and Yann Fleureau and Guillaume Lample and Stanislas Polu},
  title = {NuminaMath},
  year = {2024},
  publisher = {Numina},
  journal = {Hugging Face repository},
  howpublished = {\url{[https://huggingface.co/AI-MO/NuminaMath-1.5](https://github.com/project-numina/aimo-progress-prize/blob/main/report/numina_dataset.pdf)}}
}

@article{zhang2025spad,
  title={SPAD: Specialized Prefill and Decode Hardware for Disaggregated LLM Inference},
  author={Zhang, Hengrui and Patel, Pratyush and Ning, August and Wentzlaff, David},
  journal={arXiv preprint arXiv:2510.08544},
  year={2025}
}

@misc{digitalocean-gpu-memory-bandwidth,
  title = {GPU Memory Bandwidth and Its Impact on Performance},
  author = {DigitalOcean Community},
  year = {2023},
  url = {https://www.digitalocean.com/community/tutorials/gpu-memory-bandwidth},
  note = {Accessed: 2025-11-12}
}

@INPROCEEDINGS{bipartite_matching,
  author={Hopcroft, John E. and Karp, Richard M.},
  booktitle={12th Annual Symposium on Switching and Automata Theory (swat 1971)}, 
  title={A n5/2 algorithm for maximum matchings in bipartite}, 
  year={1971},
}

@INPROCEEDINGS{wei2023prophet,
  author={Wang, Wei and Lai, Zhiquan and Li, Shengwei and Liu, Weijie and Ge, Keshi and Liu, Yujie and Shen, Ao and Li, Dongsheng},
  booktitle={IEEE CLUSTER}, 
  title={Prophet: Fine-grained Load Balancing for Parallel Training of Large-scale MoE Models}, 
  year={2023},
}

@misc{rajbhandari2022deepspeedmoe,
      title={DeepSpeed-MoE: Advancing Mixture-of-Experts Inference and Training to Power Next-Generation AI Scale}, 
      author={Samyam Rajbhandari and Conglong Li and Zhewei Yao and Minjia Zhang and Reza Yazdani Aminabadi and Ammar Ahmad Awan and Jeff Rasley and Yuxiong He},
      year={2022},
      eprint={2201.05596},
      archivePrefix={arXiv},
      primaryClass={cs.LG},
      url={https://arxiv.org/abs/2201.05596}, 
}

@inproceedings{hwang2023tutel,
  title={Tutel: Adaptive mixture-of-experts at scale},
  author={Hwang, Changho and Cui, Wei and Xiong, Yifan and Yang, Ziyue and Liu, Ze and Hu, Han and Wang, Zilong and Salas, Rafael and Jose, Jithin and Ram, Prabhat and others},
  booktitle=MLSYS,
  year={2023}
}

@inproceedings {li2023lina,
    author = {Jiamin Li and Yimin Jiang and Yibo Zhu and Cong Wang and Hong Xu},
    title = {Accelerating Distributed {MoE} Training and Inference with Lina},
    booktitle = ATC,
    year = {2023}
}

@INPROCEEDINGS{yao2024interlayeraffinity,
  author={Yao, Jinghan and Anthony, Quentin and Shafi, Aamir and Subramoni, Hari and DK Panda, Dhabaleswar K.},
  booktitle=IPDPS, 
  title={Exploiting Inter-Layer Expert Affinity for Accelerating Mixture-of-Experts Model Inference}, 
  year={2024},
}

@inproceedings{aditya2025podattention,
    author = {Kamath, Aditya K. and Prabhu, Ramya and Mohan, Jayashree and Peter, Simon and Ramjee, Ramachandran and Panwar, Ashish},
    title = {POD-Attention: Unlocking Full Prefill-Decode Overlap for Faster LLM Inference},
    year = {2025},
    booktitle = ASPLOS
}

@inproceedings{zhu2025nanoflow,
    author = {Zhu, Kan and Gao, Yufei and Zhao, Yilong and Zhao, Liangyu and Zuo, Gefei and Gu, Yile and Xie, Dedong and Tang, Tian and Xu, Qinyu and Ye, Zihao and Kamahori, Keisuke and Lin, Chien-Yu and Wang, Ziren and Wang, Stephanie and Krishnamurthy, Arvind and Kasikci, Baris},
    title = {NanoFlow: towards optimal large language model serving throughput},
    year = {2025},
    booktitle = OSDI
}
